\documentclass[reprint, superscriptaddress, amsmath, amssymb, aps, prb]{revtex4-1} 

\usepackage{graphicx}
\usepackage{color}
\usepackage{microtype}
\usepackage{dcolumn}
\usepackage{bm}
\usepackage{hyperref}

\hypersetup{
    colorlinks   = true, 
    urlcolor     = blue, 
    linkcolor    = blue, 
    citecolor    = blue  
}

\makeatletter
\def\overbar{
  \@ifnextchar[
    {\overbar@i}{\overbar@i[1.5]}
}
\def\overbar@i[#1]{
  \@ifnextchar[
    {\overbar@ii{#1}}{\overbar@ii{#1}[1.5]}
}
\def\overbar@ii#1[#2]#3{
  \mkern #1mu\overline{\mkern-#1mu #3 \mkern-#2mu}\mkern #2mu
}
\makeatother

\def\barh{\overbar[4.7][1.5]{H}}
\def\barw{\overbar[2.5][0.8]{w}}
\DeclareMathOperator{\diag}{diag}
\def\epc{\, ,}
\def\epp{\, .}
\def\av{\mathbf{a}}
\def\bv{\mathbf{b}}
\def\cv{\mathbf{c}}
\def\dv{\mathbf{d}}
\def\ev{\mathbf{e}}
\def\nv{\mathbf{n}}

\begin{document}

\title{Anisotropic magnetic interactions and spin dynamics in the spin-chain compound\\ \texorpdfstring{Cu(py)$_2$Br$_2$}{CPB}:
An experimental and theoretical study}

\author{J.~Zeisner}
\affiliation{Leibniz Institute for Solid State and Materials Research IFW Dresden, D-01069 Dresden, Germany}
\affiliation{Institute for Solid State Physics, TU Dresden, D-01069 Dresden, Germany}

\author{M.~Brockmann}
\affiliation{Department of Physics, University of Wuppertal, D-42097 Wuppertal, Germany}

\author{S.~Zimmermann}
\affiliation{Leibniz Institute for Solid State and Materials Research IFW Dresden, D-01069 Dresden, Germany}
\affiliation{Institute for Solid State Physics, TU Dresden, D-01069 Dresden, Germany}

\author{A.~Wei{\ss}e}
\affiliation{Max Planck Institute for Mathematics, P.O.~Box 7280, D-53072 Bonn, Germany}

\author{M.~Thede}
\affiliation{Laboratory for Solid State Physics, ETH Z\"{u}rich, 8093 Z\"{u}rich, Switzerland}

\author{E.~Ressouche}
\affiliation{Universit\'{e} Grenoble Alpes, 38042 Grenoble, France}
\affiliation{MEM-MDN, INAC, 38054 Grenoble, France}

\author{K.~Yu.~Povarov}
\affiliation{Laboratory for Solid State Physics, ETH Z\"{u}rich, 8093 Z\"{u}rich, Switzerland}

\author{A.~Zheludev}
\affiliation{Laboratory for Solid State Physics, ETH Z\"{u}rich, 8093 Z\"{u}rich, Switzerland}

\author{A.~Kl{\"u}mper}
\affiliation{Department of Physics, University of Wuppertal, D-42097 Wuppertal, Germany}

\author{B.~B{\"u}chner}
\affiliation{Leibniz Institute for Solid State and Materials Research IFW Dresden, D-01069 Dresden, Germany}
\affiliation{Institute for Solid State Physics, TU Dresden, D-01069 Dresden, Germany}

\author{V.~Kataev}
\affiliation{Leibniz Institute for Solid State and Materials Research IFW Dresden, D-01069 Dresden, Germany}

\author{F.~G{\"o}hmann}
\affiliation{Department of Physics, University of Wuppertal, D-42097 Wuppertal, Germany}

\date{\today}

\begin{abstract}
We compare theoretical results for electron spin resonance (ESR) properties of the Heisenberg-Ising 
Hamiltonian with ESR experiments on the quasi-one-dimensional magnet Cu(py)$_2$Br$_2$ (CPB). Our 
measurements were performed over a wide frequency and temperature range giving insight into spin 
dynamics, spin structure, and magnetic anisotropy of this compound. By analyzing the angular 
dependence of ESR parameters (resonance shift and linewidth) at room temperature we show that 
the two weakly coupled inequivalent spin chain types inside the compound are well described by 
Heisenberg-Ising chains with their magnetic anisotropy axes perpendicular to the chain direction 
and almost perpendicular to each other. We further determine the full $g$-tensor from these data. 
In addition, the angular dependence of the linewidth at high temperatures gives us access to the 
exponent of the algebraic decay of a dynamical correlation function of the isotropic Heisenberg chain. 
From the temperature dependence of static susceptibilities we extract the strength of the exchange 
coupling ($J/k_B = 52.0\,\text{K}$) and the anisotropy parameter ($\delta\approx -0.02$) of the model Hamiltonian. An 
independent compatible value of $\delta$ is obtained by comparing the exact prediction for the resonance 
shift at low temperatures with high-frequency ESR data recorded at $4\,\text{K}$. The spin structure in 
the ordered state implied by the two (almost) perpendicular anisotropy axes is in accordance with 
the propagation vector determined from neutron scattering experiments. In addition to undoped 
samples we study the impact of partial substitution of Br by Cl ions on spin dynamics. From the 
dependence of the ESR linewidth on doping level we infer an effective decoupling of the anisotropic 
component~$J\delta$ from the isotropic exchange $J$ in these systems.
\end{abstract}

\maketitle


\section{\label{sec:introduction} Introduction}

\vspace{-0.5ex}
Although known for decades, one dimensional (1d) electronic systems
remain an active field of research in modern solid-state physics.
These systems possess their own specific phenomenology. At half
band-filling even an infinitesimal residual on-site repulsion drives
them into a Mott-insulating phase\cite{LiWu68} in which antiferromagnetic
exchange is the predominant interaction. For this reason a variety of
quasi-1d antiferromagnetic chain and ladder compounds exists in
nature. They are generally well described by the Heisenberg spin
chain with nearest-neighbor exchange or by one of its many variations
that can be obtained by coupling several chains, by extending the
range of the exchange interaction, or by making it anisotropic.
Depending on the specific choice of the exchange and anisotropy
parameters and on the strength of an applied magnetic field, these
models can have gapped or gapless excitations. In any case there are
a number of numerical and analytical methods specific for one spatial
dimension which allow for the computation of more of the
experimentally accessible quantities than for the same models in
higher dimensions. These methods include the many variants of the
numerical DMRG method\cite{DKSV04,FeWh05,Schollwoeck05,SiKl05} and
exact diagonalization\cite{Fehske_book,Weisse2013} as well as methods from
conformal\cite{LuPe75,Haldane81,BPZ84} and relativistic integrable
massive quantum field theory\cite{GNT04,Smirnov92} in 1+1 dimensions.

The variety of theoretical methods applicable to 1d~systems
boosted the search for experimental realizations of such systems
with reduced (magnetic) dimensionality starting in the seventies
of the last century (see e.g.~Ref.~\onlinecite{Mikeska2004} and
references therein). The aim of this search was, on the one hand,
to find experimental evidence for the above-mentioned physics
specific for 1d systems. On the other hand, investigations of
these materials could serve for a validation (or falsification) of
theoretical methods with potential application to higher
dimensional systems. The organo-metallic compound Cu(py)$_2$Cl$_2$
(py denotes the molecule pyridine NC$_5$H$_5$) was one of the
first realizations of a spin-1/2 Heisenberg chain and was intensively
studied some decades ago.\cite{Endoh1974,Duffy1974,Ajiro1975}
Although discovered at the same time, the closely related compound
Cu(py)$_2$Br$_2$ (CPB) received considerably less attention.
Nevertheless, as can be concluded from measurements of specific heat
and static magnetic susceptibility, CPB turned out to be closer
to an 1d material than its Cl containing counterpart.\cite{Thede2012}
Based on these measurements, it was found that CPB has an exchange
interaction along the chain not too big compared with magnetic
fields that can be realized in a laboratory, but big enough
compared to the interchain coupling.\cite{Thede2012} Thus, CPB is
a promising candidate for a 1d system suited for comparison of
experimental data with theoretical predictions.

In this work, we present such a comprehensive comparison combining
ESR as well as magnetization measurements with calculations based
on recently developed techniques. The temperature dependence of 
the magnetization enables us to determine the strength of the 
isotropic intrachain exchange ($J/k_B = 52.0\,\text{K}$) and
to estimate the value of the magnetic anisotropy ($\delta \approx -0.02$).
Results of angular dependent measurements of the ESR linewidth and 
resonance position at room temperature and at a frequency of $9.56\,\text{GHz}$ 
can be explained considering the existence of two magnetically 
inequivalent chains in this material as well as a small anisotropy 
$\delta$. Furthermore, based on these measurements we determine the 
$g$-tensor of this compound and find evidence for the presence of 
two anisotropy axes, related to the different types of chains. 
A possible spin configuration of the 
ordered state, which follows from this structure, is compatible 
with the propagation vector $(0, 0.5, 0.5)$ obtained from neutron 
scattering investigations. From frequency dependent 
high-field/high-frequency ESR (HF-ESR) measurements we derive
the temperature independent value of the $g$-factor along the chain
axis $g_c = 2.153$. The experimentally determined $g_c$ allows us to calculate 
the resonance shift of the ESR line from HF-ESR data measured at 
$4\,\text{K}$. By comparing the obtained resonance shifts with 
shifts calculated by means of field theoretical and exact methods, 
we show that exact finite temperature calculations (or at least 
logarithmic corrections to field theory) are required in order to 
describe the low-temperature data. Finally, we discuss ESR studies 
on samples with two different amounts of partial substitution of Br 
by Cl ions. From the change of the linewidth with doping 
concentration we conclude an effective decoupling of anisotropic 
exchange from isotropic exchange as function of doping.

The paper is organized as follows. In Sec.~\ref{sec:general_remarks}
we recall part of the theoretical background for the exact
calculation of the thermodynamics of the Heisenberg chain and for
the description of microwave absorption probed in ESR experiments.
Sec.~\ref{sec:experimental_methods} is devoted to details of the
samples, the methods and the equipment used in our experiments. In
Sec.~\ref{sec:magnetization} we explain how the anisotropy can be
extracted from two magnetization measurements with magnetic fields
applied in two different directions. The analysis of our ESR
experiments is presented in Sec.~\ref{sec:ESR_CPB}. 
Sec.~\ref{sec:neutron_scattering} accounts for the results of 
neutron scattering experiments on CPB. In Sec.~\ref{sec:doping}
we discuss the influence of substituting 
a small amount of the Br by Cl ions. Finally, in Secs.~\ref{sec:discussion}
and \ref{sec:conclusions}, we discuss our results and conclude by summarizing 
the main statements of the paper and by giving an outlook to 
possible future studies. In the appendices we present two new 
theoretical methods used in this work, one for analyzing 
magnetization data of close-to-isotropic models (App.~\ref{app:susceptibility}), 
another one for analyzing line shift and linewidth of the
resonance lines (ESR parameters) by means of (modified) 
moments of the spectral function (App.~\ref{app:ESR_parameters}). 
In App.~\ref{app:spin_structure} we discuss the spin structure of 
the ordered ground state of CPB using a renormalization group 
argument.

\section{\label{sec:general_remarks} Theoretical background}

From the analysis of our thermodynamic and ESR measurements we
shall argue that the magnetic properties of the compound CPB are
well described by the spin-1/2 Heisenberg-Ising chain (or XXZ chain)
\begin{equation}\label{eq:XXZ-model}
  \mathcal{H} = J \sum_{\langle ij \rangle}
                     \left[\bm s_i \cdot \bm s_j +\delta \, s_i^zs_j^z\right]
\end{equation}
with exchange interaction of strength $J$ and anisotropy parameter
$\delta$. More precisely, our experimental data can 
be interpreted consistently, for temperatures down to $4\,\text{K}$, 
assuming that the two inequivalent magnetic chains inside the 
compound are described by two non-interacting XXZ chains with the 
same values of $J$ and $\delta$ but two different orientations of 
the magnetic symmetry axes (called `the anisotropy axes' in
the following). In doing so, we neglect weak interchain 
couplings which lead to a 3d ordering temperature of about 
$T_N=0.72\,\text{K}$.\cite{Thede2012}

The Hamiltonian \eqref{eq:XXZ-model} defines one of the most
studied and best understood 1d many-particle models. It belongs to
the class of so-called integrable lattice models,\cite{thebook} meaning
that, in addition to the generic 1d methods mentioned in the previous
section, several advanced mathematical techniques can be applied
to calculate its thermodynamic properties\cite{Kluemper04,Takahashi99}
and some of its thermal correlation functions\cite{GKS04a,BDGKSW08}
analytically. For the comparison with our magnetization measurements
we shall resort to the so-called quantum transfer matrix approach to
the thermodynamics of integrable lattice models.\cite{Kluemper1992,%
Kluemper1993} This approach allows us to calculate the magnetization
and the neighbor-correlation functions, that are needed to take into 
account a small anisotropy, exactly and to arbitrary precision for 
the Heisenberg model on an infinite chain.

The correlation function which determines the absorption of microwaves
in ESR experiments within linear response theory\cite{Kubo1954}
and which is therefore relevant for our work is the imaginary part
of the dynamical susceptibility,
\begin{equation}\label{eq:dynamical_susceptibility}
  \chi_{+-}''(\omega,h) = \frac{1}{2L} \int_{-\infty}^\infty {\rm d} t \:
                     {\rm e}^{{\rm i} \omega t}
                     \left\langle [S^+ (t), S^-] \right\rangle_{T, h,\delta} \,.
\end{equation}
Here, $L$ is the number of lattice sites in the spin chain, $S^+$
and $S^-$ are ladder operators for the total spin, and the
brackets under the integral denote the thermal average in the
canonical ensemble at temperature $T$ and for an external magnetic
field of strength $H$ with corresponding Zeeman energy $h=g\mu_B\mu_0 H$. 
The direction of the magnetic field is, in our convention, the $z$ 
direction. For later convenience, we include the parameter $\delta$ 
of Hamiltonian \eqref{eq:XXZ-model} into the list of subscripts of 
the thermal average. In App.~\ref{app:ESR_parameters} we discuss 
more general set-ups where, for instance, the incident wave is linearly 
polarized rather than circularly, as well as a slightly more general 
Hamiltonian whose anisotropy axis is arbitrarily oriented.

The ESR line is determined by the absorbed intensity
$I(\omega,h) = \omega \chi''(\omega,h)/2$. In spite of the
integrability of the XXZ chain an analytic calculation of this
function at all temperatures and magnetic fields is still out of
reach. Numerical calculations based on the exact diagonalization
of finite chains \cite{Ikeuchi2017,Brockmann2011,Brockmann2012}
are plagued by finite size effects, rendering them unreliable for
small temperatures and small anisotropies. Small anisotropies
cause narrow absorption lines, meaning that a high numerical
frequency resolution is required or, alternatively, that we
need to know the corresponding time-dependent correlation
functions in the long-time limit. As far as we understand,
this also restricts the applicability of current finite-temperature
dynamical DMRG methods. Field theoretical methods,\cite{Oshikawa1999,Oshikawa2002}
on the other hand, are suitable for small anisotropies, but are
restricted to small temperatures and a limited range of
magnetic fields. 

Instead of calculating the full dynamical
susceptibility, one may try to find appropriate measures for
certain characteristic features of the spectral line, like the
deviation of its center from the position of the paramagnetic
resonance, the so-called resonance shift, or its linewidth (for
details see App.~\ref{app:ESR_parameters}). Such an approach was
originally proposed by van Vleck \cite{VanVleck1948}, who devised
a `method of moments' even before the linear response theory was
invented. Van Vleck found formulae for the moments in the 
high-temperature limit. Later, Maeda et al.~\cite{MSO05} related
the resonance shift of the XXZ chain with small anisotropy to a
certain nearest-neighbor static correlation function which can
be extracted from the free energy per lattice site and can be
computed exactly for arbitrary temperatures and magnetic fields.
In previous work\cite{Brockmann2011,Brockmann2012} part of the
authors developed a general method of moments for the XXZ
model in an external magnetic field directed along the magnetic
anisotropy axis. It relates all moments of the normalized
intensity $I(\omega,h)/I_0$ to static finite-range correlation
functions. In 1d the first few of them can be exactly calculated
for arbitrary temperature, magnetic field, and anisotropy.\cite{BDGKSW08,TGK10a}
They provide an idea about the temperature and field dependence of 
the ESR parameters. The question if this dependence can be observed 
experimentally stood at the beginning of our work.

In the comparison of moment-based ESR parameters with
experimental data from standard ESR experiments, two possible
difficulties may arise. The first one relates to the fact that the
moments are calculated as integrals over the frequency for fixed
magnetic field, while ESR experiments are usually performed for
fixed frequency and the field is varied. As we have pointed out in
previous work\cite{Brockmann2012} this may even cause a seemingly
wrong prediction for the qualitative behavior of the linewidth as a
function of temperature. Still, the discrepancy can be resolved,
at least in principle, by changing the experimental set-up such
that the frequency is varied at fixed external field. In practice,
however, such a frequency sweep measurement with fixed magnetic
field is rather challenging (see e.g.~Ref.~\onlinecite{Wiemann2015} 
and references therein), in particular, when dealing with broad 
resonance lines. 

A second difficulty which may be encountered is that the linewidth
defined by the second moment of the absorbed intensity may take
rather different values than its width at half height, which is
one of the standard experimental measures of the linewidth. The
reason is that `long tails' of the resonance line may considerably
contribute to the moment-based linewidth while they are entirely
ignored by a measure like the width at half height. In the
experimental ESR data such tails may be overlaid by background
noise which makes an extraction of the moment-based width from the
data problematic if not impossible. In this work we try to
overcome this problem by introducing moments in which the absorbed
intensity is multiplied by a `weight function' providing a cut-off
for the high-frequency tails (see App.~\ref{app:generalized_moments}). 
For small anisotropy and high temperatures a scaling analysis then 
makes it possible to relate the moment-based width with the width 
at half height. This way we can understand and interpret the angular 
dependence of our high-temperature data for the linewidth of CPB. 
Our interpretation supports the picture of `inhibited exchange 
narrowing' developed in Ref.~\onlinecite{Hennessy1973}.

\section{\label{sec:experimental_methods} Samples and experimental methods}

\begin{figure}[t]
  \includegraphics[width=\columnwidth]{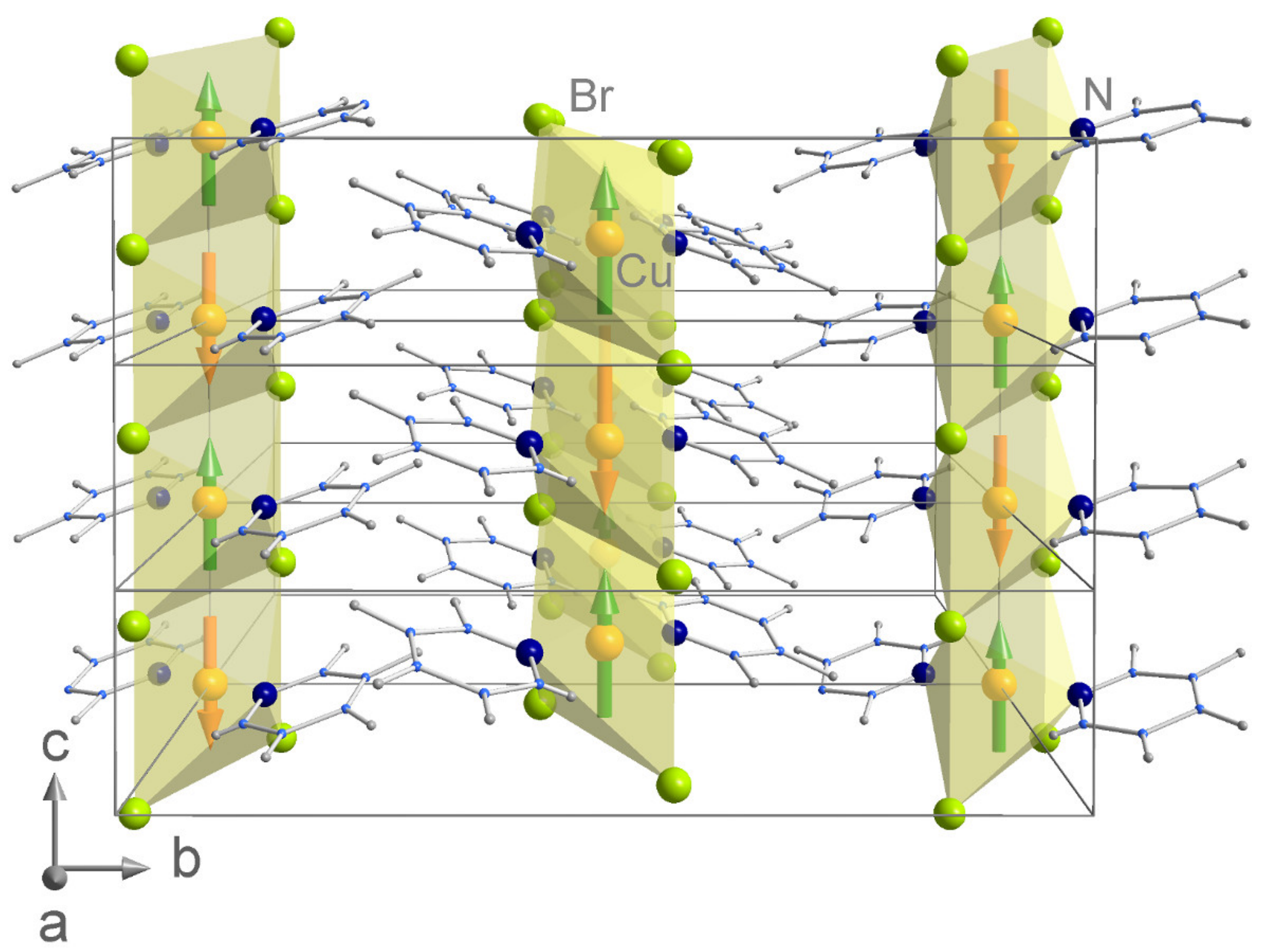}
    \caption{Structure of the compound Cu(py)$_2$Br$_2$. Cu ions
    (yellow), located in the centers of stretched octahedra (Br green, N dark blue), 
    form chains along the $c$ axis which are separated from each 
    other by pyridine rings NC$_5$H$_5$ (C light blue, H gray). 
    Crystallographic data are taken from Ref.~\onlinecite{Morosin1975}. 
    The arrows indicate the proposed magnetic structure of CPB 
    below $T_{N}\simeq 0.72\,\text{K}$ as discussed in 
    Sec.~\ref{sec:neutron_scattering} and App.~\ref{app:spin_structure}.}
    \label{fig:CPB_structure}
\end{figure}

\begin{figure}[t]
    \includegraphics[width=\columnwidth]{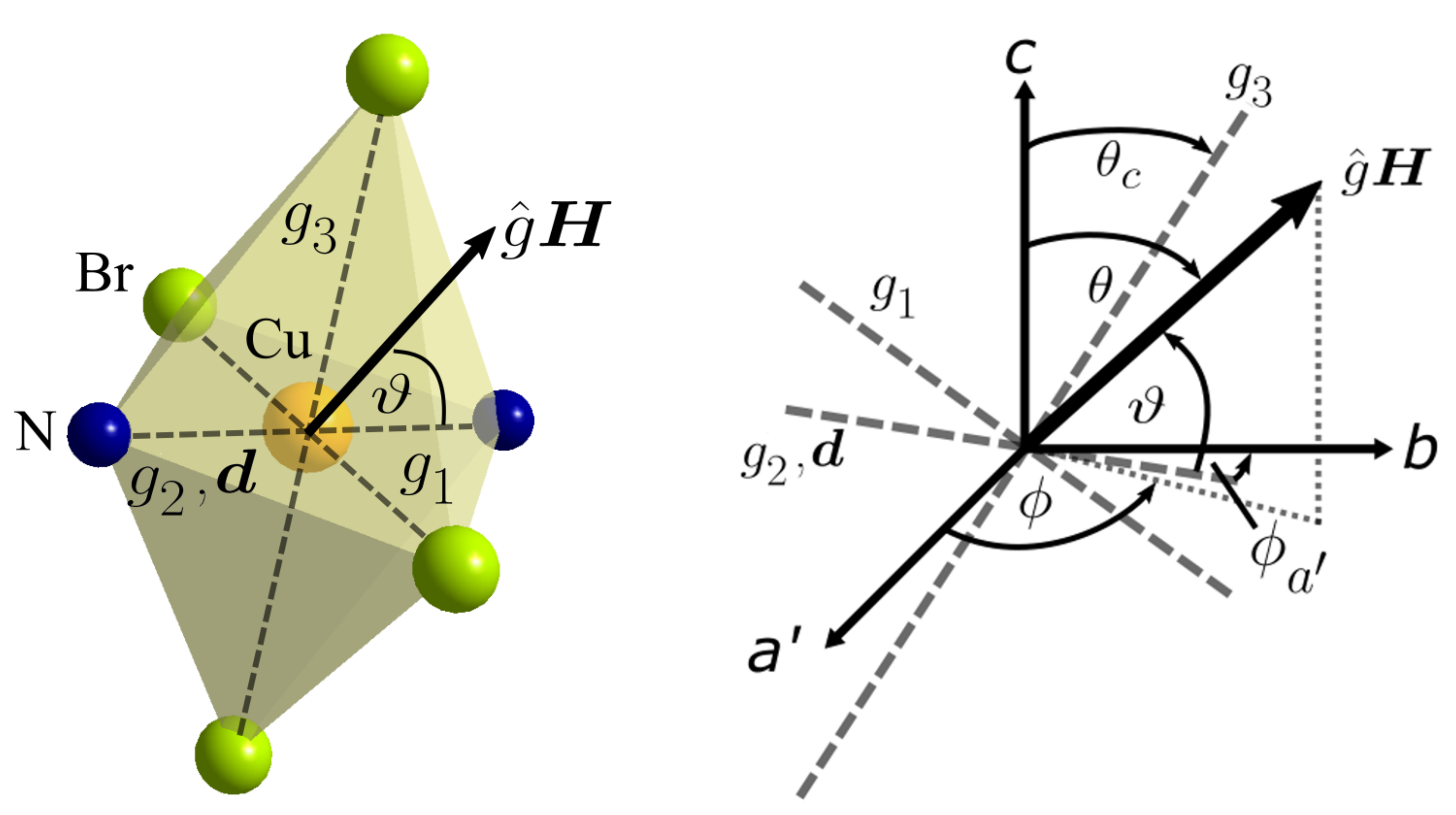}
    \caption{Left: Local coordinate system of a stretched octahedron 
    formed by four bromine ions (green) and two nitrogen ions (dark 
    blue), surrounding the central copper ion (yellow). Principal 
    axes $g_1$, $g_2$, and $g_3$ of the $g$-tensor $\hat{g}$
    coincide with the symmetry axes of the stretched octahedron. The angle
    between local magnetic field $\hat{g}\bm H$ and anisotropy
    axis $\dv$ is denoted by $\vartheta$. Right: Angles of $\hat{g}\bm H$ ($\theta$, $\phi$), $\dv$ ($90^\circ, 90^\circ-\phi_{a'}$), 
    and $g_3$ ($\theta_c, -\phi_{a'}$) with respect to the crystallographic frame $(\av',\bv,\cv)$. 
    Additionally, the angle $\vartheta$ between $\hat{g}\bm H$ and $\dv$ is shown.}
    \label{fig:coordinate_systems}
\end{figure}

Single crystals used in this study were grown from solution and
were investigated by means of measurements of static
susceptibility, specific heat and muon spin rotation in Ref.~\onlinecite{Thede2012}.
A crucial input for the discussion of our ESR data below is the
crystallographic structure of our samples. CPB is monoclinic
($P2_1/m$) with $a = 8.424\,\text{\AA}$, $b = 17.599\,\text{\AA}$,
$c = 4.0504\,\text{\AA}$, and $\beta = 97.12^{\circ}$.\cite{Morosin1975} 
The magnetic ions are Cu$^{2+}$ ions ($S = 1/2$) which form
chains along the $c$ axis (see Fig.~\ref{fig:CPB_structure}).
Each of these Cu ions is surrounded by four Br ligands and two N
ligands, the latter belonging to the pyridine molecules which
separate neighboring chains from each other. The surrounding
ligands form a stretched octahedron whose stretching axis, i.e.~the
longer Br-Cu-Br axis, is tilted away from the $c$ axis by an
angle $\theta_c = 37.24^{\circ}$, as shown in Fig.~\ref{fig:coordinate_systems}.
The angle between the projection of the stretching axis onto the
plane perpendicular to the $c$ axis (called $a'$-$b$ plane in the
following) and the $a'$ axis is $\pm\phi_{a'}$ with $\phi_{a'} = 43.44^{\circ}$
for the two inequivalent chains. The line connecting the two
opposite nitrogen ligands almost lies in the $a'$-$b$ plane,
tilted away only by $0.3^\circ$. It encloses an angle of
$\pm(90^\circ - \phi_{a'}) = \pm 46.56^\circ$ with the $a'$ axis.
Single crystals cleave along the $c$ axis, which enables us to
easily identify this crystallographic direction.

There are two magnetically inequivalent types of chains which
differ in the orientation of the stretching axis of the octahedra.
They can be transformed into each other by combining a
reflection with respect to a plane normal to the $b$ axis
lying in between the two chains and a translation of
$\boldsymbol{c}/2$ in $c$ direction (see Fig.~\ref{fig:CPB_structure}).
Therefore, the orientation of the ionic $g$-tensors is different
for these two chain types, while the $g$-tensors for sites
within one chain are identical.

Neighboring magnetic ions in the individual chains are
antiferromagnetically coupled by superexchange via the halogen 
ligands between them. The strength of this intrachain exchange was
obtained in Ref.~\onlinecite{Thede2012} by comparing the static
susceptibility measured in a field along the chain direction with
the exact result for the isotropic Heisenberg chain,\cite{Johnston2000}
given by Hamiltonian \eqref{eq:XXZ-model} with $\delta=0$. The
authors of Ref.~\onlinecite{Thede2012} report an isotropic
exchange of $J = 4.58\,\text{meV}$. Although neighboring chains are 
well separated from each other, there exists a residual interchain 
exchange $J'$ which leads to 3d ordering at finite temperatures. This 
transition was observed\cite{Thede2012} in specific heat measurements 
at $T_N=0.72\,\text{K}$ and can be used to estimate the strength of 
the interchain exchange to be $J' \approx 0.03\,\text{meV}$ (see 
e.g.~Ref.~\onlinecite{Yasuda2005}). From these values it follows 
that the magnetic interactions in CPB have a strong one-dimensional 
character thus qualifying this compound for comparison with theories 
based on 1d models like Eq.~\eqref{eq:XXZ-model}.

We measured static magnetization of a CPB sample using a VSM-SQUID
magnetometer from Quantum Design Inc.~in DC-mode in the temperature
range from $1.8\,\text{K}$ to $325\,\text{K}$, in order to
reinvestigate the exchange coupling $J$ by taking the effect of a
small anisotropy $\delta$ into account.

Beside the pure compound CPB, two doped samples with 2\% and 5\% Cl
content were studied. Their crystal structure is similar to CPB
with some of the Br sites occupied by Cl ions, which leads to local
changes of the $g$-tensor and of the effective isotropic exchange.\cite{Thede2012} 
This way disorder is introduced into the system.

For our ESR studies of these compounds two spectro\-meters were
employed. Measurements with a microwave frequency of
$9.56\,\text{GHz}$ at temperatures between $3.6\,\text{K}$ and  $300\,\text{K}$, and fields up to $0.9\,\text{T}$ were performed
using a standard Bruker EMX X-Band spectro\-meter.
HF-ESR was measured using a
homemade spectrometer which is described in detail elsewhere.\cite{Golze2006} All HF-ESR measurements were
performed in transmission geometry and Faraday configuration,
i.e.~with wave vector of the microwaves being parallel to the
external field.

The neutron diffraction measurements were performed on D23
instrument in Institut Laue-Langevin (\mbox{Grenoble}, France). The fully
deuterated sample of CPB was mounted on the dilution refrigerator
stick, installed on a standard ILL Orange cryostat. Incident neutron
beam with wavelength $\lambda=2.375\,\text{\AA}$ was provided by the PG
monochromator. The measurements were performed in a standard
geometry with a single $^3$He detector.

\section{\label{sec:magnetization} Magnetization}

\begin{figure}[t]
  \includegraphics[width=0.99\columnwidth]{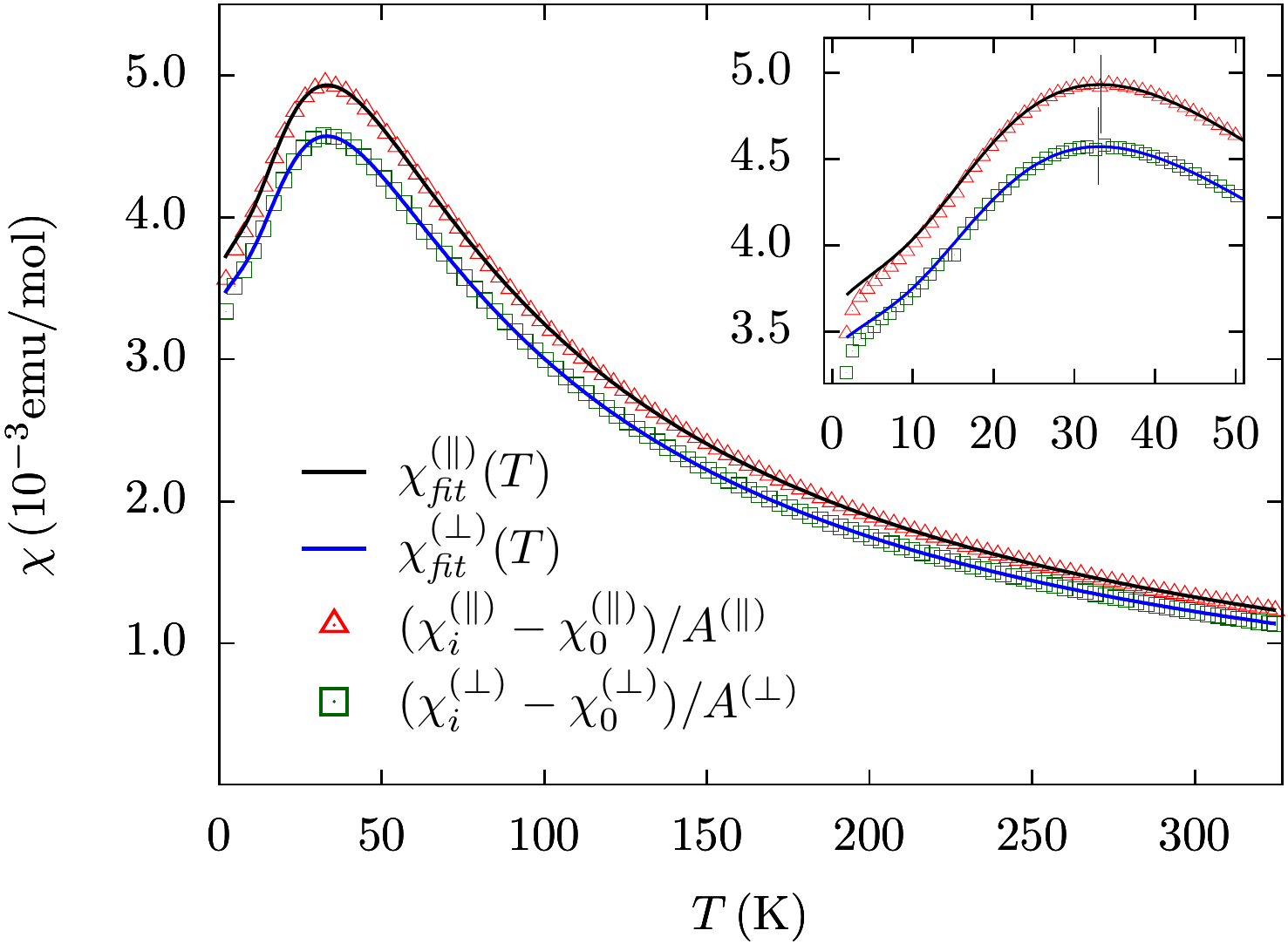}
  \caption{Static susceptibilities of CPB for two orientations of 
  the small magnetic field of 0.1~T ($\|$ and $\perp$ to the $c$ axis) 
  as functions of temperature. Open symbols indicate measured
  data (minus offsets and divided by geometry factors, 
  see Eqs.~\eqref{eq:chi_perturbation}). For the sake of clarity, 
  only every sixth of these data points is plotted in the main plot and every 
  second point in the inset. Solid lines show the best fit
  $\chi_\text{\textit{fit}}^{(\|,\perp)}(T) = \chi^{(0)}(T) + \chi_\text{corr}^{(\|,\perp)}(T)$,
  corresponding to $J/k_B = 52.0\,\text{K}$ and $\delta = -0.019$ (see text). 
  The excellent match between calculated and measured data is 
  emphasized in the inset. Vertical lines indicate the positions of 
  the Bonner-Fisher maxima.  The difference in height can be mainly attributed to 
  the $g$-factors of the two field directions.}
  \label{fig:CPB_susc}
\end{figure}

The temperature dependence of the magnetization of a CPB sample
was measured with a small applied field of about $0.1\,\text{T}$ upon heating after
zero field cooling. In one of the measurements the external field was
oriented approximately along the chain axis, while in another one
it was applied nearly perpendicular to this axis. In the following
we neglect small misalignments of the crystal and consider
susceptibilities defined as the magnetization divided by the small
field of $0.1\,\text{T}$ (see App.~\ref{app:susceptibility}).
We label the two susceptibilities and the corresponding data sets
by $(\|)$ for $\boldsymbol{H}\, \|\, [001]$ and by $(\perp)$ for
$\boldsymbol{H}\, \perp\, [001]$, respectively. Static
susceptibilities extracted from the two measurements are shown in
Fig.~\ref{fig:CPB_susc}. 

For both orientations the behavior of the
susceptibility is qualitatively similar, showing a Bonner-Fisher
maximum,\cite{Bonner1964} which is typical for spin-1/2 chains and
whose position and height are mainly related to the strength of the
exchange interaction. The fact that the two susceptibility curves 
differ from each other by a constant factor over a wide temperature 
range can be mainly attributed to the $g$-factor anisotropy, which 
can be extracted from the angular dependence of the resonance field 
of our ESR data at high temperatures (see Sec.~\ref{sec:angular_dependence}), 
and to geometry factors taking the sample shape into account. The 
small difference of the positions of the two maxima can be 
explained by a small anisotropy of the exchange interaction. 
Assuming the former to be of Ising type we may use first order 
perturbation theory (see App.~\ref{app:susceptibility}), valid for 
all temperatures $T \gg J\delta/k_B$ with Boltzmann's constant $k_B$, 
in order to estimate the parameter $\delta$ of Eq.~\eqref{eq:XXZ-model}.

From the angular dependence of the ESR data in
Sec.~\ref{sec:angular_dependence} we conclude that the anisotropy
axes of the spin chains in our material are perpendicular to the
$c$ axis. This means that for $\chi^{(\|)}$ the magnetic field is
perpendicular to the anisotropy axes. Denoting the magnetic field direction by $z$,
the perturbation term becomes $J\delta \sum_{\langle ij \rangle} s_i^x s_j^x$,
and the first order correction to the isotropic susceptibility,
$\chi^{(0)}(T) = g^2\mu_B^2\left\langle s_1^z \right\rangle_{T,h,0}/h$
with Zeeman energy $h = g\mu_B\mu_0 H$, takes the form (see
App.~\ref{app:susceptibility_example})
\begin{equation}\label{eq:chi1_corr}
  \chi^{(\|)}_\text{corr}(T) =
     \frac{g^2\mu_B^2 J \delta}{h}
     \frac{\rm d}{{\rm d}h} \left\langle s_1^x s_2^x \right\rangle_{T,h,0}\,.
\end{equation}
Here, the subscripts at $\left\langle\cdot\right\rangle_{T,h,0}$
mean that the thermal expectation value has to be evaluated with
the isotropic Hamiltonian, i.e.~Eq.~\eqref{eq:XXZ-model} with
$\delta = 0$, supplemented by the Zeeman term $-hS^z = -g\mu_B \mu_0 H \sum_j s_j^z$. 

For $\chi^{(\perp)}$ the magnetic field lies in the $a'$-$b$ 
plane. Denoting its direction again by $z$, the anisotropic part 
of the Hamiltonian of one of the two inequivalent chains in CPB 
reads
\begin{equation}\label{eq:Ham_theta}
  \mathcal{H}_{\vartheta} =
     J\delta \sum_{\langle ij \rangle}
        (\cos\vartheta\, s_i^z - \sin\vartheta\, s_i^x)
    (\cos\vartheta\, s_j^z-\sin\vartheta\, s_j^x)\,,
\end{equation}
where $\vartheta$ is the angle between magnetic field and the
corresponding anisotropy axis. If we take into account that the
anisotropy axes of the two chains are almost perpendicular to each
other, and if we further neglect the small anisotropy of the 
$g$-factor inside the $a'$-$b$ plane (see Sec.~\ref{sec:angular_dependence}), 
the first order contribution of both chain types to the total 
susceptibility simplifies to the arithmetic mean of the individual 
contributions and is therefore given by
\begin{equation}\label{eq:chi2_corr}
  \chi^{(\perp)}_\text{corr}(T)
     = \frac{g^2\mu_B^2 J \delta}{2h}
       \frac{\rm d}{{\rm d}h} \left\langle s_1^z s_2^z + s_1^x s_2^x \right\rangle_{T,h,0}\,.
\end{equation}
Everything is now reduced to quantities that can be calculated
exactly in the thermodynamic limit. The isotropic part
$\chi^{(0)}(T)$ of the static susceptibility and its corrections
\eqref{eq:chi1_corr} and \eqref{eq:chi2_corr} can be most efficiently
computed by solving a simple and finite set of non-linear integral
equations arising within the so-called quantum transfer matrix
approach to the thermodynamics of integrable lattice models.\cite{Kluemper1992,Kluemper1993}

We fitted the theoretical predictions
\begin{subequations}\label{eq:chi_perturbation}
\begin{align}
  \chi^{(\|)}(T) &= A^{(\|)}\left(\chi^{(0)}(T) + \chi^{(\|)}_\text{corr}(T)\right) + \chi_0^{(\|)} \,, \label{eq:chi1_perturbation}\\
  \chi^{(\perp)}(T) &= A^{(\perp)}\left(\chi^{(0)}(T) + \chi^{(\perp)}_\text{corr}(T)\right) + \chi_0^{(\perp)} \,, \label{eq:chi2_perturbation}
\end{align}
\end{subequations}
to the measured data $ \chi_i^{(\|)}$ and $\chi_i^{(\perp)}$,
respectively. Here, $A^{(\|,\perp)}$ are dimensionless geometry
factors and $\chi_0^{(\|,\perp)}$ are offsets of the data sets
$\chi_i^{(\|,\perp)}$ measured in units $\text{emu/mol}$. We derive
the general structure of these equations in App.~\ref{app:susceptibility_general}.

The fit values of the isotropic coupling $J$ and the anisotropy
parameter $\delta$ depend on the chosen temperature range $[T_a,T_b]$
of the fit. We varied the lower bound $T_a$ from $16\,\text{K}$
to $32\,\text{K}$ and the upper bound $T_b$ from $200\,\text{K}$
to $325\,\text{K}$. Values of $T_\text{a}$ smaller than
$22\,\text{K}$ or values of $T_b$ larger than $285\,\text{K}$
suddenly decrease the quality of the fit. The former makes sense
since a perturbation expansion, as given in Eqs.~\eqref{eq:chi_perturbation},
is only valid for $T\gg J\delta/k_B \approx 1\,\text{K}$. The
latter is due to more noise and perhaps a systematic error in the
susceptibility data above room temperature. The best fit is obtained
for $T_a=23.5\,\text{K}$ and $T_b=285\,\text{K}$ and yields
\begin{align}\label{eq:J_from_chi_fit}
  J/k_B &= 52.0\,\text{K} \pm 0.1\,\text{K}\,, \\[0.5ex]
  \delta &= -0.019 \pm 0.002 \approx -0.02\,.\label{eq:delta_from_chi_fit}
\end{align}
Offsets and prefactors are
$\chi_0^{(\|)} = 1.46\cdot 10^{-4}\, \text{emu}/\text{mol}$,
$\chi_0^{(\perp)} = -2.77 \cdot 10^{-4}\,\text{emu}/\text{mol}$
and $A^{(\|)}\cdot(g^{(\|)})^2 = 4.48$,
$A^{(\perp)}\cdot(g^{(\perp)})^2 = 4.55$, respectively. If we set
$g^{(\|)} = g_c = 2.154$ and $g^{(\perp)} \approx 2.069$ as obtained
by means of ESR spectroscopy in Sec.~\ref{sec:ESR_CPB}, the latter
value being an estimated average over $g$-values in the $a'$-$b$
plane, both geometry factors are close to one, $A^{(\|)} = 0.97$,
$A^{(\perp)} = 1.06$.

Figure~\ref{fig:CPB_susc} shows the data sets
$(\chi_i^{(\|,\perp)}-\chi^{(\|,\perp)}_0)/A^{(\|,\perp)}$ together
with the two curves
$\chi^{(0)}(T)+\chi_\text{corr}^{(\|,\perp)}(T)$ of the best fit
with $\delta = -0.019$ and $J/k_B = 52.0\,\text{K}$.  
The fit provides a reliable estimate of the anisotropy in CPB (for 
details see App.~\ref{app:susceptibility}). The relative positions
of the two maxima (see inset of Fig.~\ref{fig:CPB_susc}) already 
give a clear hint at the sign of~$\delta$. The fact that the 
position of the maximum of $\chi^{(\|)}(T)$ is slightly shifted to 
higher temperatures as compared to the one of $\chi^{(\perp)}(T)$ 
implies that the anisotropy is negative and small (see Eq.~\eqref{eq:delta_from_chi_CPB} 
in App.~\ref{app:susceptibility_example}) meaning that the
Hamiltonian is critical in zero magnetic field.

In the low-temperature regime $T\leq 3\,\text{K}$ our
susceptibility data show a strong decrease with decreasing
temperatures and the curves obtained from a perturbation expansion
in $\delta$ deviate from the experimental data (see inset of
Fig.~\ref{fig:CPB_susc}). This is compatible with the fact that the
perturbation expansion is only valid for $T\gg J\delta/k_B \approx 1\,\text{K}$.
The low-temperature behavior might be qualitatively explained by
an effective magnetic excitation gap which opens if the applied
field is perpendicular to the anisotropy axis\cite{Hieida2001,Dmitriev2002}
or by the proximity of the 3d antiferromagnetic phase transition
at $T_N \approx 0.72\,\text{K}$.

\section{\label{sec:ESR_CPB} ESR on \texorpdfstring{Cu(py)$_2$Br$_2$}{CPB}}

\subsection{\label{sec:angular_dependence} Angular dependence of ESR parameters}

We study the angular dependence of the ESR spectrum of CPB at
room temperature and for a fixed frequency of $\nu=9.56\,\text{GHz}$.
We recorded three data sets. For two of them the $c$ axis of the
sample was initially aligned with the external field, then rotated
away from the field direction by $90^\circ$. A third data set
pertains to a rotation about the $c$ axis which enclosed an angle
of $90^\circ$ with the external field. This data set corresponds to
a rotation of the field in the $a'$-$b$ plane in the reference
frame of the sample. In Fig.~\ref{fig:coordinate_systems} we show
the local octahedral environment of the Cu$^{2+}$ ions of one of
the two inequivalent chains and its relative position to the
crystallographic frame $(\av',\bv,\cv)$.

\begin{figure}[t]
	\includegraphics[width=\columnwidth]{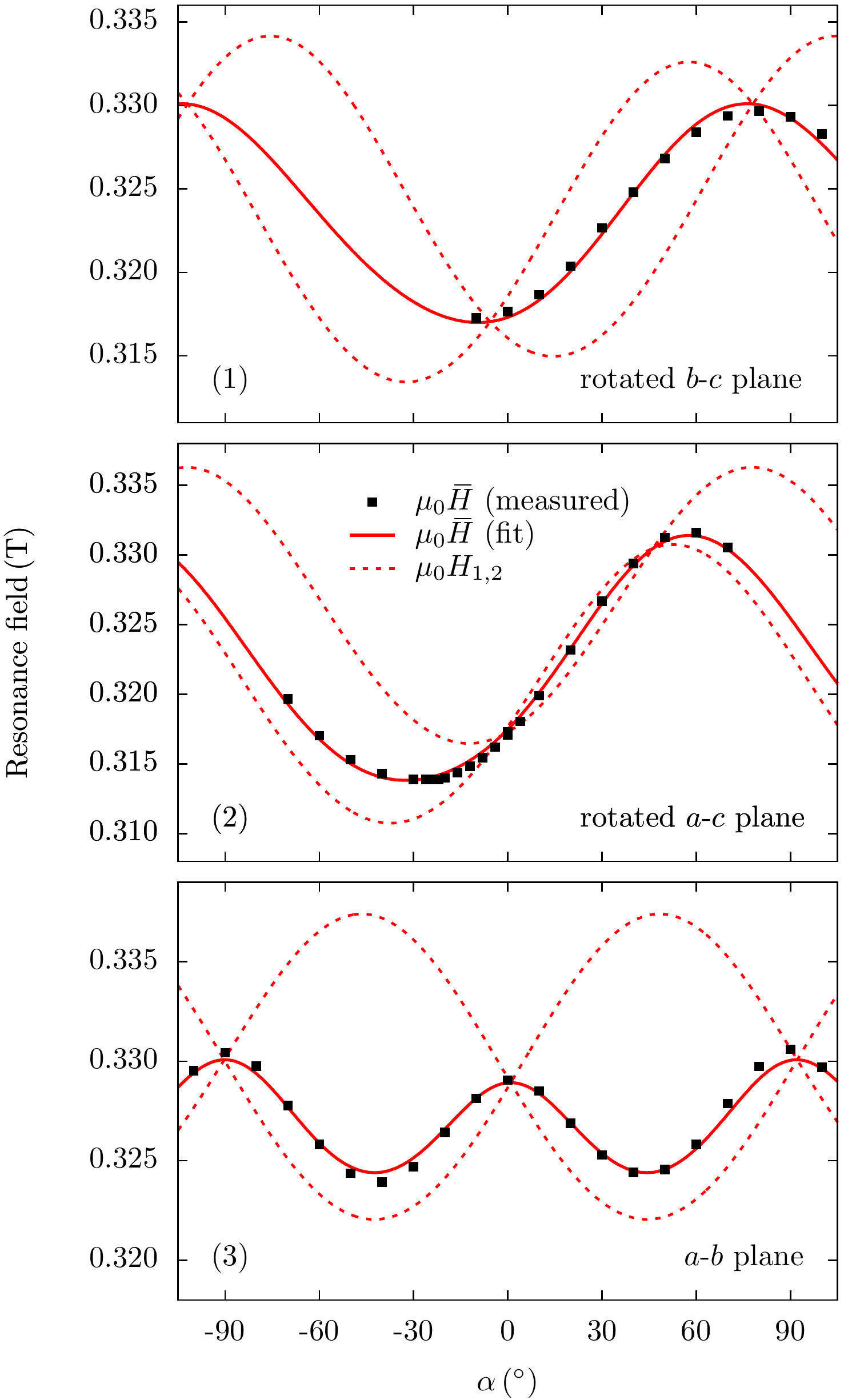}
	\caption{Angular dependence of the measured resonance field for CPB
		(dots) at frequency $\nu=9.56\,\text{GHz}$ and
		at room temperature for rotation of the magnetic field in planes
		containing the $c$ axis (top and middle) and in the $a'$-$b$
		plane (bottom), compared to the fitted theoretical curves
		$\mu_0 \protect\barh$ (solid lines). Dashed lines indicate the resonance fields $\mu_0 H_{1,2}$ of the two individual, unresolved lines (see Eq.~\eqref{eq:Hbar} and text).}
	\label{fig:X-band_ang_dep_I}
\end{figure}

\begin{figure}[t]
	\includegraphics[width=\columnwidth]{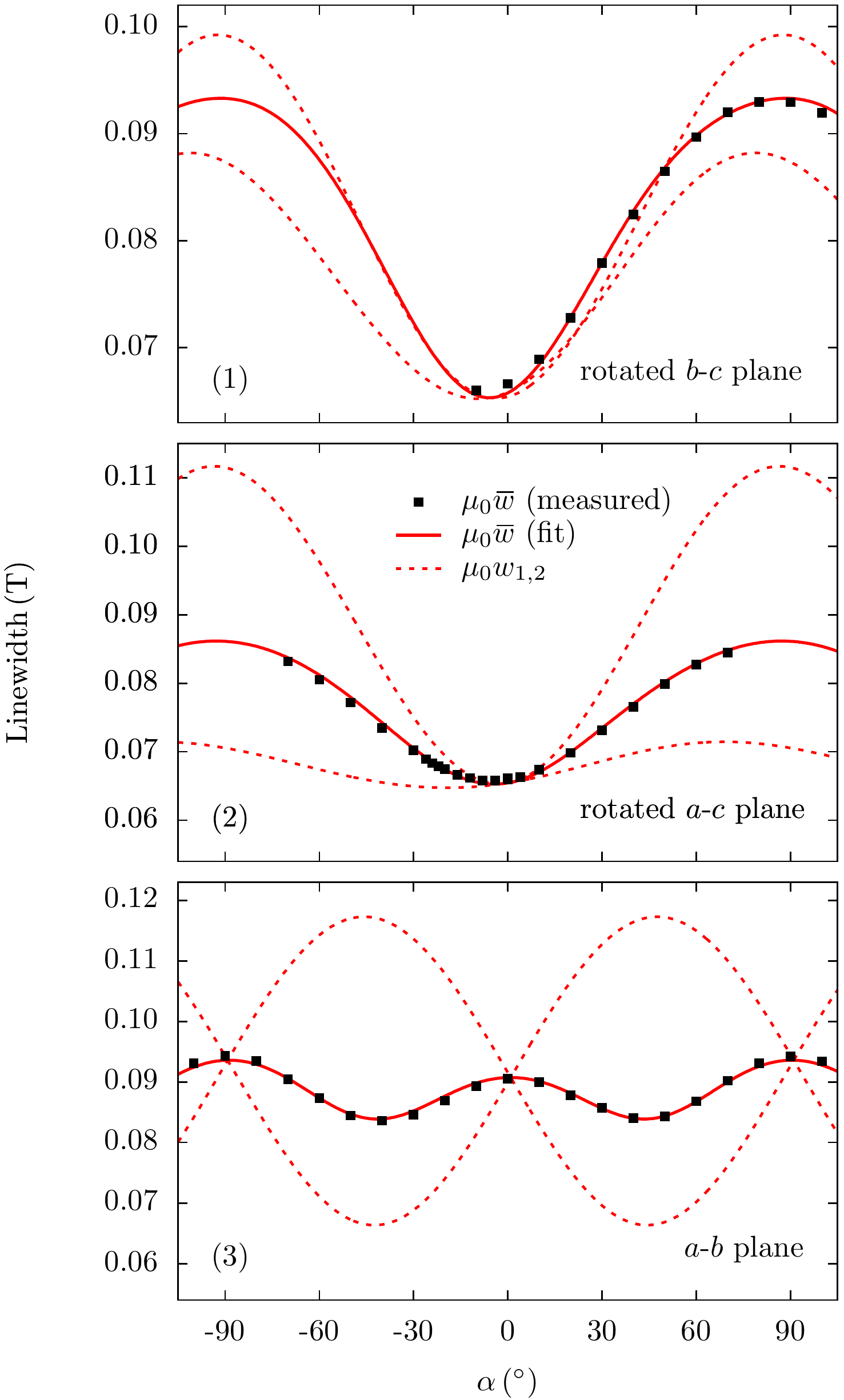}
	\caption{Angular dependence of the measured ESR linewidth for CPB (dots) at frequency $\nu=9.56\,\text{GHz}$ and at room
		temperature for rotation of the magnetic field in planes
		containing the $c$ axis (top and middle) and in the $a'$-$b$
		plane (bottom), compared to the fitted theoretical curves
		$\mu_0 \protect\barw$ (solid lines). Dashed lines indicate the linewidths $\mu_0 w_1$ and $\mu_0 w_2$
		of the two individual, unresolved lines (see Eq.~\eqref{eq:wbar} and text).}
	\label{fig:X-band_ang_dep_II}
\end{figure}

All recorded spectra show single spectral lines from which we
extracted the resonance fields $\barh$ and linewidths $\barw$ as
functions of the rotation angle $\alpha$. The corresponding curves
of ESR parameters are shown in Figs.~\ref{fig:X-band_ang_dep_I}
and \ref{fig:X-band_ang_dep_II} as black squares.

It turns out that the analysis of these curves is rather intricate.
This is first of all due to the fact that we are dealing with two
inequivalent chains (see Fig.~\ref{fig:CPB_structure}) meaning that
we have to interpret the recorded spectral lines as superpositions
of two individual lines which are so close to each other that they
are not resolved at the applied frequency. Note that in principle
---besides the exchange narrowing effect due to intrachain
interaction given by $J$--- there might be an additional exchange
narrowing effect caused by interchain interactions $J'$, which would
lead to the fusion of the two spectral lines.\cite{Pal1994, Hennessy1973}
Such effect might be anticipated from the fact that in case of CPB $J'$ is 
larger than the difference in Zeeman energies of the inequivalent chains. 
However, angular dependencies of ESR parameters of the resulting single line 
are not compatible with our results obtained by rotation of magnetic field 
in a-b-plane, which can be described only in terms of contributions from two 
individual lines, see below. 

A second difficulty arising in the analysis of our data comes from
the fact that the octahedra surrounding the magnetically active
Cu$^{2+}$ ions are distorted in such a way that the cubic symmetry
of the undistorted octahedra is fully broken. This implies that we
are dealing with the most general possible g-tensor which, as a
symmetric rank two tensor, depends on six independent parameters,
e.g.~its eigenvalues $g_1, g_2, g_3$ and three angles fixing its
orientation in space. We may therefore write it as
\begin{equation} \label{eq:gform}
     \hat g = D \diag (g_1,g_2,g_3) D^t \epc
\end{equation}
where $D$ is the rotation matrix transforming the principle
coordinate system of the $g$-tensor into the crystallographic
frame $(\av',\bv,\cv)$. The $g$-factor anisotropy is caused by
spin-orbit coupling which mixes, in the case of Cu$^{2+}$ ions,
some of the $t_{2g}$ states to the ground state, i.e.~the
$d_{x^2-y^2}$ state.\cite{Abragam_Bleaney} Since the anisotropy is
small we expect a close-to-isotropic $g$-tensor,
i.e.~$g_1 \approx g_2 \approx g_3$.


We analyze the angular dependence of the ESR parameters
based on the model of two non-interacting inequivalent XXZ chains. 
In the following, we denote Bohr's magneton by $\mu_B$, the
permeability of free space by $\mu_0$, and Planck's constant by
$2\pi\hbar$. The letter $h$ is already used as abbreviation for
the Zeeman energy $h=g\mu_B\mu_0 H$ and should not be confused
with Planck's constant. For a single chain our theory relies on perturbation theory in
$\delta$, on an analysis of the moments of the shape function,
and on a high-temperature expansion in $J/(k_B T)$ for the resonance
shift $s(\delta)$ and the linewidth $w$ (see App.~\ref{app:generalized_moments}).
The resonance shift is related to the resonance field $H$ by
\begin{equation}\label{eq:res_cond}
     H = \frac{\hbar\omega - J s(\delta)}{g\mu_B\mu_0 } \epp
\end{equation}
Here, $\nu = \omega/(2\pi)$ is the frequency of the incident
microwaves and $g = \|\hat g \ev\|$, where $\ev$ is the unit vector
in the direction of the external magnetic field. To leading order
in $J/(k_B T)$ we obtain the following expression for the resonance
shift (see Eq.~\eqref{eq:HTE_m1_tilde} of App.~\ref{app:generalized_moments}),
\begin{equation}\label{eq:HTE_shift}
     s(\delta) = \frac{J\delta}{4k_B T}\left[(1-3\cos^2\vartheta)\frac{\hbar\omega}{J}
          + (1+\cos^2\vartheta)\frac{\delta}{2}\right]
\end{equation}
with $\vartheta$ being the angle between the magnetic field
direction $\hat g \ev/g$ at the Cu sites and the anisotropy axis of
the chain. Note that up to first order in $\delta$ the frequency
term $\hbar\omega$ can be replaced by $g\mu_B\mu_0 H$. This
relation will be also proven useful for the analysis of
high-frequency ESR measurements at high temperatures in the next
section.

In 1d systems the usual exchange narrowing argument fails. It can
be replaced by a modified argument, leading to `inhibited exchange
narrowing'.\cite{Hennessy1973} Further elaborating on this idea we
derive a novel formula for the linewidth for small $\delta$
and in the high-temperature regime (see App.~\ref{app:generalized_moments}),
\begin{equation}\label{eq:HTE_linewidth}
     w \approx \frac{AJ}{g\mu_B\mu_0}
       \left[\frac{\delta^2}{4}(1+\cos^2\vartheta)\right]^\beta \epp
\end{equation}
The proportionality factor $A$ is unknown and should be of order
one. As explained in App.~\ref{app:generalized_moments}, the exponent
$\beta$ is connected with the decay of a certain time-dependent
correlation function in the isotropic system at high temperature.

Eqs.~\eqref{eq:res_cond}, \eqref{eq:HTE_shift} and \eqref{eq:HTE_linewidth}
determine the resonance field and linewidth of the absorption
spectrum of a single XXZ chain with small anisotropy and in the
high-temperature regime. We still have to take into account that the
observed spectra must be interpreted as the superposition of the
spectra of two types of chains, type~1 and type~2, which are
distinguished by the orientation of their $g$-tensors and
anisotropy axes. We shall assume for simplicity that in its center
each of the two spectral lines can be approximated by a Lorentzian
and that the two lines have equal spectral weight. For two equally
normalized Lorentzians with maxima at $H_1$, $H_2$ and widths $w_1$,
$w_2$ their sum is well approximated again by a Lorentzian if only
$|H_1 - H_2| \ll \text{min}\{w_1,w_2\}$. The location of the maximum 
of the resulting line is approximated by
\begin{equation}\label{eq:Hbar}
     \barh = \frac{H_1 w_1^{-3} + H_2 w_2^{-3}}{w_1^{-3} + w_2^{-3}}
               + \mathcal{O}(\epsilon_H^3 \epsilon_w)
\end{equation}
and its width $\barw$ can be expressed as
\begin{align}
     \barw & = \sqrt{ \sqrt{w_1^2 w_2^2 + \frac{(w_1-w_2)^4}{4}}-\frac{(w_1-w_2)^2}{2} }
                 \notag\\ & \quad\,
         + \frac{(H_1-H_2)^2}{w_1+w_2}
           \left(\frac{3}{4}
              -\frac{25}{8}\left(\frac{w_1-w_2}{w_1+w_2}\right)^2\right)
         \notag\\ & \quad\,
         + \mathcal{O}(\epsilon_H^3,\epsilon_H^2\epsilon_w^4) \label{eq:wbar}
\end{align}
with small numbers $\epsilon_w = (w_1-w_2)/(w_1+w_2)$ and
$\epsilon_H = (H_1-H_2)/(w_1+w_2)$. The formula for the location
$\barh$ of the maximum represents a weighted mean of the two 
resonance fields $H_1$ and $H_2$ with weights $1/w_{1,2}^3$. Note 
that it holds for other line shapes than Lorentzians, e.g.~for a 
superposition of two Gaussians, too. The first line in expression~\eqref{eq:wbar}
for the resulting width $\barw$ can be understood as a modified
geometric mean of two individual widths $w_1$ and $w_2$, whereas
the second line reflects an additional broadening caused by the
finite distance of the maxima positions.

We fit derivatives of Lorentzians to the measured spectral
lines as, due to the use of lock-in techniques, the derivative of
the absorption line was recorded in our low-frequency ESR
experiments. We identify Lorentz parameters `position' and `width'
with $\barh$ and $\barw$ of Eqs.~\eqref{eq:Hbar}
and \eqref{eq:wbar}, respectively. For the individual resonance
shifts $H_1$, $H_2$ and linewidths $w_1$, $w_2$ of the two types
of lines we have used Eqs.~\eqref{eq:res_cond} and \eqref{eq:HTE_linewidth}
with the respective orientations of the $g$-tensors and anisotropy
axes. Taking these equations as they are, the number of parameters
to be determined would be too large for a stable fit. Ideally
the following parameters of the model should be extracted
from a fit: the anisotropy $\delta$, the eigenvalues $g_1$, $g_2$,
$g_3$ of the $g$-tensor, $2 \times 3$ angles fixing the rotation
matrices $D_1$ and $D_2$ that determine the orientation of the
$g$-tensors of the two types of chains, $2 \times 2$ angles
fixing two unit vectors $\dv_1$, $\dv_2$ defining the
direction of the anisotropy axes of the two types of chains, and
finally the parameters $A$ and $\beta$ entering Eq.~\eqref{eq:HTE_linewidth}.

In order to reduce the number of unknowns of the fit, we fix
the `geometric parameters' $D_1$, $D_2$ and $\dv_1$, $\dv_2$ by
resorting to the crystal structure (see Sec.~\ref{sec:experimental_methods})
and by inspecting the qualitative behavior of the data. We have seen
in Sec.~\ref{sec:experimental_methods} that the two inequivalent
chains in CPB are related by a glide reflection with reflection
component $R = \diag (1,-1,1)$ representing a reflection at the
$a$-$c$ plane. This implies that $R D_1 = D_2 R$ and $R \dv_1 =
\dv_2$, i.e.~$g$-tensors and anisotropy axes of the two chains
must be related by this reflection. It is convenient to specify the
direction $\ev$ of the external magnetic field in terms of spherical
coordinates $\theta$, $\phi$ with respect to the crystallographic
frame $(\av',\bv,\cv)$. Then, $\ev = \ev (\theta, \phi)$ and
the $g$-factors $g_j = \|\hat g_j \ev\|$, $j = 1, 2$, of the two
chains become functions of $\theta$ and $\phi$. Eq.~\eqref{eq:gform}
implies that $g_j (\theta,\phi)$ is periodic in $\theta$ with
period $180^\circ$, that $g_j (0,\phi)$ is periodic in $\phi$,
also with period $180^\circ$, and that $g_2(\theta,\phi) = g_1(\theta,-\phi)$.

The most striking feature of the experimental resonance shift and
linewidth shown in Figs.~\ref{fig:X-band_ang_dep_I} and \ref{fig:X-band_ang_dep_II}
is that they exhibit a $180^\circ$ periodicity if the field is
rotated in planes perpendicular to the $a'$-$b$ plane, but a
$90^\circ$ periodicity if the field is rotated within the $a'$-$b$
plane. The $g$-factors of the individual chains have a periodicity
of $180^\circ$ for all rotation directions. The periods of
resonance field and linewidth induced by the anisotropy of the
individual chains are $180^\circ$, too, as can be seen from
Eqs.~\eqref{eq:HTE_shift}, \eqref{eq:HTE_linewidth}. Thus, any
shorter period or modulation must come from the superposition of
the resonance lines of the two chains.

Let us first consider the variation of the linewidth (see Eqs.~\eqref{eq:HTE_linewidth}
and \eqref{eq:wbar} as well as Fig.~\ref{fig:X-band_ang_dep_II}).
In Eq.~\eqref{eq:HTE_linewidth} the variation of the $g$-factor
with the external field is a subleading effect, the main variation
of the width coming from the variation of $\vartheta$. In the upper
two panels of Fig.~\ref{fig:X-band_ang_dep_II} no modulation of the
$180^\circ$ periodicity is visible, showing that both angles
$\vartheta_1$ and $\vartheta_2$ and thus both individual widths
$w_1$ and $w_2$ have the same monotonic behavior as function of
rotation angle $\alpha$. By contrast, the $90^\circ$ modulation of
the width in the lower panel points towards a phase difference of
about $90^\circ$ between $\vartheta_1$ and $\vartheta_2$. This can
be understood if the anisotropy axes lie in the $a'$-$b$ plane and
are almost perpendicular to each other. Taking into account that
$R \dv_1 = \dv_2$, they should enclose an angle of about $45^\circ$
with the $a'$ axis. Thus, the anisotropy axis should either be
directed along the projection of the stretched octahedron axes onto
the $a'$-$b$ plane or perpendicular to this direction. Only the
latter case is (approximately) in accordance with the reflection
symmetries of the deformed octahedra. For this reason we conclude
that the anisotropy axes of the chains are located in the $a'$-$b$
plane and enclose angles $\pm (90^\circ - \phi_{a'}) = \pm 46.56^\circ$
with the $a'$ axis. As we shall see this will also explain the
behavior of the resonance field, Eqs.~\eqref{eq:res_cond} and
\eqref{eq:HTE_shift}, if the $g$-tensor anisotropy is properly
taken into account.

For the $g$-tensor anisotropy we hypothesize that it is entirely
due to the deformation of the octahedra formed by the Br and N
atoms surrounding the magnetically active Cu$^{2+}$ spin. Then,
the $g$-tensor should be diagonal in a coordinate system
symmetrically attached to the deformed octahedra. Denoting by
$D(\alpha,\nv)$ the matrix for a rotation about an axis $\nv$ by
an angle $\alpha$, we are setting
\begin{subequations}
\label{eq:gtensor_orientation}
\begin{align}
     D_1 &= D( \phi_{a'}, \cv) D(-\theta_c, \bv)\,, \\[0.3ex]
     D_2 &= D(-\phi_{a'}, \cv) D(-\theta_c, \bv)\,,
\end{align}
\end{subequations}
which means that we are neglecting the small declination away from
the $a'$-$b$ plane of the line connecting the nitrogen atoms in
the octahedron (see Sec.~\ref{sec:experimental_methods} and
Fig.~\ref{fig:coordinate_systems}). The above notation is also
useful to represent $\dv_1$ and $\dv_2$ explicitly as
\begin{equation} \label{eq:anisotropy_orientation}
     \dv_{1, 2} = D(\pm \phi_{a'}, \cv) \bv = D(\pm(90^\circ-\phi_{a'}), \cv) \av' \,,
\end{equation}
which means that the anisotropy axis of each chain coincides with
the connecting line of the two nitrogen ions (see Fig.~\ref{fig:coordinate_systems}).

Presuming Eqs.~\eqref{eq:gtensor_orientation} and \eqref{eq:anisotropy_orientation}
we have reduced the model parameters to be fitted to the angular
dependence of the high-temperature ESR data to $\delta$,
$g_1$, $g_2$, $g_3$, $A$, and $\beta$. We reduce the number of
these parameters further by using $\delta = -0.019$ as obtained
from our susceptibility measurements. Except for these model
parameters we also have to determine some experimental parameters
connected with the limited control over the sample position
during our measurements, which are described below.

The best fit yields for the remaining model parameters
\begin{align}
  (g_1,g_2,g_3) &= (2.065,\,2.018,\,2.203)\,, \label{eq:g_fit}\\[0.3ex]
  A &= 1.3\,, \label{eq:A_fit}\\[0.3ex]
  \beta &=0.77\,. \label{eq:beta_fit}
\end{align}
The estimated error of $\beta$ is about $3\,\%$ and those of the
three $g$-values $g_{1,2,3}$ are less than $0.5\,\%$. The three
values of $g_{1,2,3}$ in Eq.~\eqref{eq:g_fit} together with
Eqs.~\eqref{eq:gtensor_orientation} determine the full $g$-tensor
for both chain types and are typical for Cu$^{2+}$ ions in an
octahedral environment.\cite{Abragam_Bleaney} For a magnetic
field applied along the chain axes, the $g$-factors of both chain
types are the same due to reflection symmetry and take the value
$g_c = 2.154$. This value is in excellent agreement with the value
obtained independently from high-frequency measurements (see
Sec.~\ref{sec:frequency_dependence} below).

For a different choice of the model parameter $\delta$, say
$\delta = -0.01$, $-0.03$, or $-0.05$, the best fit yields
similar values of $g_1$ and $g_3$ as well as of $\beta$, all
lying in the estimated error intervals. This can be understood by
the observation that the effect of $\delta$ on the resonance
position at high temperatures in Eq.~\eqref{eq:res_cond} is very
small: $s(\delta) \sim \delta/T$. The variation of the fit
parameter $g_2$ with $\delta$ is slightly larger (up to $1.5\,\%$),
leading to values $g_2 \leq 2$ for $\delta < -0.04$. Furthermore,
the model parameter $\delta$ enters the formula of the linewidth,
Eq.~\eqref{eq:HTE_linewidth}, via the prefactor $A\cdot\delta^{2\beta}$.
If $\delta$ was too small in absolute value this would yield
values of $A$ not of order one, in contradiction to our expectation
(see App.~\ref{app:generalized_moments}). This way and by demanding
that $g_2 > 2$ we can exclude values of $\delta$ greater than $-0.01$
and less than $-0.04$, in agreement with our previous findings.

Except for the model parameters the fit yields a number of
experimental parameters, for instance `off-plane' angles $\phi_\text{op}^{(1,2)}$
and $\theta_\text{op}^{(a'b)}$. The former are angles between the
$b$-$c$ plane (label 1) or the $a$-$c$ plane (label 2) and planes
rotated about the $c$ axis by $\phi_\text{op}^{(1,2)}$. Their
meaning is that during the corresponding measurement (labels (1)
and (2) in Figs.~\ref{fig:X-band_ang_dep_I} and \ref{fig:X-band_ang_dep_II})
the crystal was rotated such that the external magnetic field was
lying in these rotated planes rather than in the unrotated $b$-$c$
or $a$-$c$ planes. During the rotation of the third measurement
(label (3) in Figs.~\ref{fig:X-band_ang_dep_I} and \ref{fig:X-band_ang_dep_II})
the $a'$-$b$ plane enclosed an angle $\theta_\text{op}^{(a'b)}$
with the external magnetic field. Since $\theta_\text{op}^{(a'b)}$
is small (see below) we can neglect it and call this a rotation of
the magnetic field inside the $a'$-$b$ plane. Further, offset
angles $\alpha_\text{os}^{(1,2,a'b)}$ are determined by the fit.
They describe (small) misalignments of the external magnetic field
with crystallographic axes, e.g.~the $c$ axis for measurements (1)
and (2) or the $a'$ axis for measurement (3), at $\alpha=0$. They
read
\begin{align}
     (\phi_\text{op}^{(1)}, \phi_\text{op}^{(2)}, \theta_\text{op}^{(a'b)})
        &= (-6.4^\circ,\,26.9^\circ,\,1.2^\circ)\,, \\
     (\alpha_\text{os}^{(1)}, \alpha_\text{os}^{(2)}, \alpha_\text{os}^{(a'b)})
        &= (5.6^\circ,\,1.9^\circ,\,-0.9^\circ)\,.
\end{align}
The values of $\theta_\text{op}^{(a'b)}$, $\alpha_\text{os}^{(2)}$,
and $\alpha_\text{os}^{(a'b)}$ are negligible. The order of
magnitude of $\alpha_\text{os}^{(1)} \sim 5^\circ - 10^\circ$ could
have been already estimated by eye from the corresponding data sets
of the upper panels of Figs.~\ref{fig:X-band_ang_dep_I} and 
\ref{fig:X-band_ang_dep_II}.

Figures~\ref{fig:X-band_ang_dep_I} and \ref{fig:X-band_ang_dep_II}
show the experimental data (black dots) together with the fitted
theoretical curves (red solid lines) for the angular dependence
of resonance position and linewidth. The red dashed lines represent
the contributions from the two inequivalent chains. The linewidth,
measured as width at half height, of the sum of a broad and a
narrow line is dominated by the width of the narrow line.
In all cases we assume equal intensities of the two lines
composing the observed spectral line. Therefore, the width of the
observed line is minimal if the linewidth of one of the two
contributing lines has a minimum (see lower panel of Fig.~\ref{fig:X-band_ang_dep_II}).
From our point of view the agreement of the fitted curves $\barh(\alpha)$
and $\barw(\alpha)$ with the measured data points is rather
convincing in all three cases.

In conclusion, from the angular dependence of the ESR parameters
measured at room temperature $T \gg J/k_B \approx 52.0\,\text{K}$,
the eigenvalues of the $g$-tensor could be determined. The scenario
of two anisotropy axes in the $a'$-$b$ plane explains the observed
angular dependence of resonance field and linewidth. Furthermore,
from heuristic arguments the possible value of $\delta$ could be
restricted to the interval $[-0.04,-0.01]$ which is compatible
with the value of $\delta = -0.019$ obtained from susceptibility
measurements. Additionally, the values of $A$ and $\beta$ in
Eq.~\eqref{eq:HTE_linewidth} could be estimated. We expect that,
due to further progress in theory, they may be calculated one day.
For the time being they provide experimentally measured quantities
of certain time-dependent correlation functions of the isotropic
Heisenberg chain. The value of $\beta = 0.77$, for instance,
is related to the algebraic long-time decay of the
finite-temperature correlation function ($T \approx 6J/k_B$) that
appears under the integral of Eq.~\eqref{eq:m2_tilde_perp} (see 
Eqs.~\eqref{eq:g_infty_asymp}-\eqref{eq:width_eta} in 
App.~\ref{app:ESR_parameters}, valid at high temperatures).

In the infinite temperature limit this correlation function
simplifies to
\begin{align}\label{eq:g_infty}
  g_\infty(t) &= \frac{4}{2^L L}\sum_{j,k=1}^L \text{Tr}\Big\{e^{iH_\text{xxx}t}s_j^+s_{j+1}^+ e^{-iH_\text{xxx}t} s_k^-s_{k+1}^-\Big\} \notag\\
  &\simeq \alpha (Jt)^{-\gamma_\infty}
\end{align}
with $\gamma_\infty = 0.6$, i.e.~$\beta_{\infty} = 0.71$ (see App.~\ref{app:generalized_moment_HTE}).
The value of $\beta = 0.77$, i.e.~$\gamma=0.70$, at $T\approx 6J/k_B$
is in accordance with a numerical analysis that we performed for
finite temperatures, $1 \leq k_B T/J \leq 100$, and up to $28$
lattice sites, similar to the one in App.~\ref{app:generalized_moment_HTE}
for infinite temperature (see e.g.~Fig.~\ref{fig:g_infty}).

\subsection{\label{sec:frequency_dependence} Frequency dependence of the resonance position}

\begin{figure}[t]
  \includegraphics[width=\columnwidth]{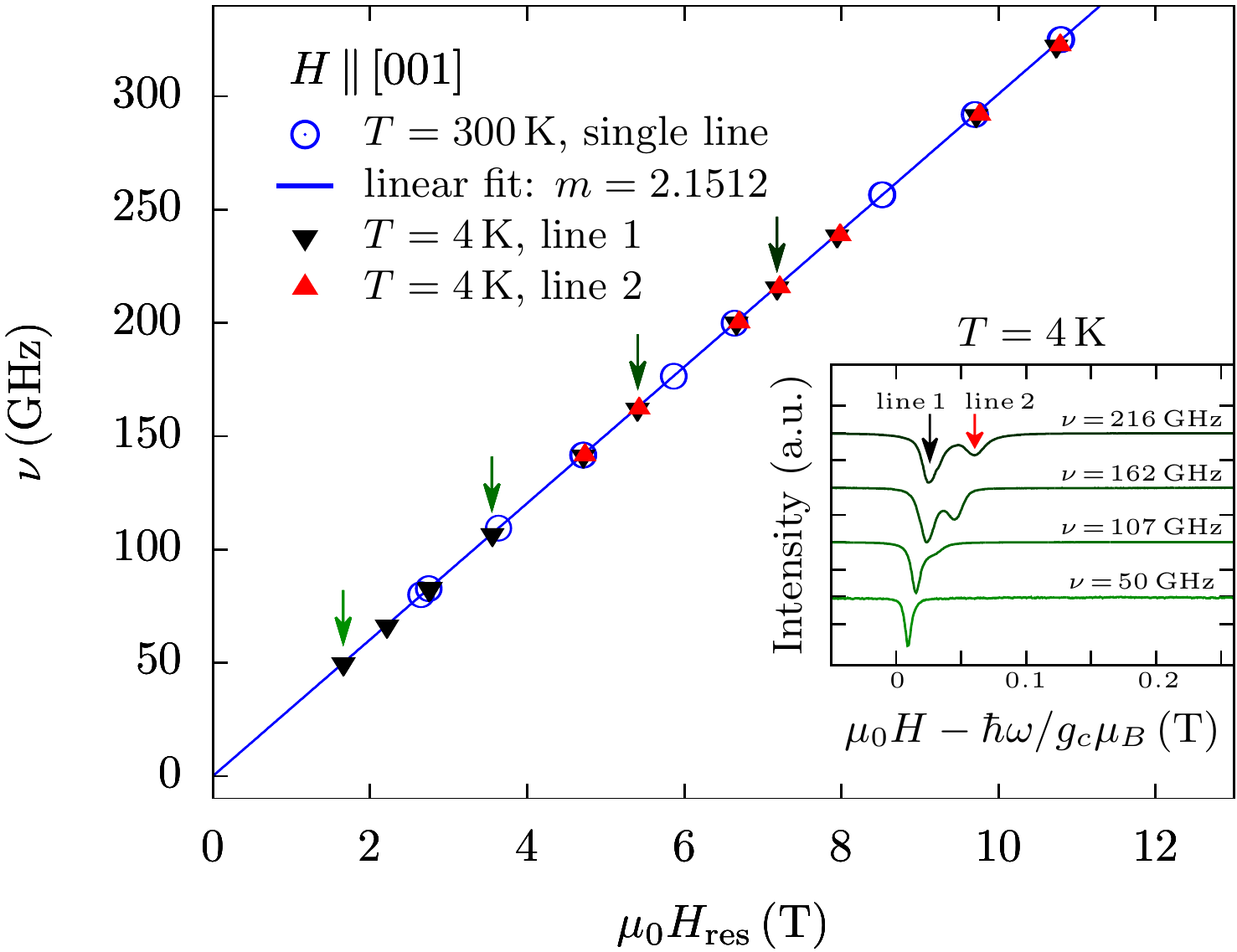}
  \caption{Frequency-field dependence of resonance positions for CPB at 
  $300\,\text{K}$ and $4\,\text{K}$ with external magnetic field 
  oriented along the $c$ axis. Arrows in the main plot indicate 
  the resonance positions of the selected ESR spectra shown in the 
  inset, recorded at $4\,\text{K}$ and, for a better comparison, 
  shifted horizontally by the paramagnetic resonance frequency as 
  well as vertically by arbitrary amounts.}
  \label{fig:CPB_HF-ESR_exp}
\end{figure}

We conducted HF-ESR studies of the resonance shift of
the spectral line for comparison with calculations presented in
Refs.~\onlinecite{Oshikawa2002}, \onlinecite{MSO05} and
Refs.~\onlinecite{Brockmann2011}, \onlinecite{Brockmann2012}.
Measurements of the frequency dependence of the ESR parameters
were performed at $4\,\text{K}$ and $300\,\text{K}$ in a frequency
range from $50\,\text{GHz}$ to $325\,\text{GHz}$ on a sample which
was oriented such that $\boldsymbol{H}\,||\,[001]$. The results
for the resonance positions at both temperatures are shown in
Fig.~\ref{fig:CPB_HF-ESR_exp}.

In paramagnets the resonance field and the absorption frequency of 
spins probed in ESR experiments are linear functions of each other. 
In the presence of spin-orbit coupling the resonating spin is
sensitive to the crystal field of its paramagnetic environment
whose reaction to an external magnetic field is then encoded in
the (ionic) $g$-tensor. Antiferromagnetic exchange coupling
between neighboring spins induces an additional shift of the
resonance position, which is a pure many-body effect and depends
on the exchange anisotropy, quantified by $\delta$ in our case.
In theory it is easy and natural to distinguish between the effect
of the $g$-tensor and the (many-body) resonance shift $s(\delta)$,
see Eq.~\eqref{eq:res_cond}. In experiments, however, it may
be difficult to separate the two effects, because $s(\delta)$
depends linearly on the field for small fields. In
Refs.~\onlinecite{Brockmann2011, Brockmann2012} some of us derived a
formula that allows to compute the resonance shift at arbitrary
temperature for a single XXZ chain with the magnetic field applied
in the direction of the anisotropy axis. A strong
deviation from the linear behavior for large enough magnetic
fields ($h/J \gtrsim 0.1$) and not too small anisotropy
(e.g.~$\delta \approx -0.1$) of the model Hamiltonian \eqref{eq:XXZ-model} was found.
However, it turned out that the anisotropy of CPB is too small and
that the magnetic fields realizable in our experiments are
not strong enough to find a pronounced deviation from the linear
behavior.

Still, a careful analysis of our data allows us to extract the
resonance shift at high and low temperatures. From the analysis of
the ESR data recorded at high-temperature and with an external
field in $c$ direction we obtain, based on Eq.~\eqref{eq:HTE_shift}
with $\vartheta = 90^\circ$, an estimate of the $g$-value $g_c$.
The shift at high temperatures is small and proportional to the
resonance field itself. Therefore, it can be absorbed into the
proportionality factor denoted by $m$ in Eq.~\eqref{eq:m_fit},
which is sometimes called an `effective $g$-factor'. The
temperature independent value $g_c$ can then be obtained by
fitting a straight line to the resonance position measured at high
temperatures, and taking the first order high-temperature
correction into account. Furthermore, higher corrections imply a
way to estimate the magnitude of $\delta$.

At low temperature the resonance shift as a function of the
resonance field shows stronger deviation from linear behavior.
Fitting different theoretical predictions\cite{Oshikawa2002,
MSO05,Brockmann2011,Brockmann2012} we shall obtain two more
estimates of the anisotropy parameter $\delta$. Both of them
are compatible with our previously obtained values within the
estimated errors.

Another approach\cite{Psaroudaki2014} that works for small system
sizes at zero temperature is based on Bethe ansatz techniques and
identifies a certain excited state above the ground state that
contributes most (as compared to all other states, at least for
small system sizes) to the ESR absorption spectrum. We computed
the difference of the energy of this state to the ground state
energy for different magnetic fields up to system size $L=256$ by
means of Bethe ansatz. The dependence of this energy difference
on the magnetic field agrees well with the corresponding resonance
shifts of the measured spectra at low temperatures for all
used frequencies and is in accordance with field theory and the
moment-based approach considered in more detail below.

\subsubsection{\label{sec:HF-ESR_HT} High temperatures}

At $T=300\,\text{K}$ we observed single resonance lines which
show a linear frequency-field relation (see Fig.~\ref{fig:CPB_HF-ESR_exp}).
In the high-temperature regime and for small anisotropies, the
resonance condition for the frequency $\nu=\omega/(2\pi)$ of the
incident microwave and the resonance field $H_\text{res}$ reads
(see Eqs.~\eqref{eq:res_cond} and \eqref{eq:HTE_shift} with $\vartheta=90^\circ$)
\begin{equation}\label{eq:res_cond_2}
   \frac{\hbar\omega}{J} =  g_c\left(1 + \frac{J\delta}{4k_B T}\right)
      \frac{\mu_B\mu_0 H_{\text{res}}}{J} + \frac{J\delta^2}{8k_B T}\,.
\end{equation}
This explicit expression is deduced from a perturbation expansion
in $\delta$ up to second order and a high-temperature expansion up
to $J/(k_B T)$ of the shifted moment $m_1$, cf.~Eqs.~\eqref{eq:averagef},
\eqref{eq:m1_tilde} and \eqref{eq:HTE_m1_tilde}. We fit the
function $y = mx+b$ with dimensionless quantities $y = \hbar\omega/J$
and $x=\mu_B\mu_0 H_{\text{res}}/J$ to the high-temperature data
and obtain
\begin{align}
  m &= g_c\left(1 + \frac{J\delta}{4k_B T}\right) = 2.1512\,, \label{eq:m_fit}\\
  b &= \frac{J\delta^2}{8k_B T} = 5 \cdot 10^{-6}\,.\label{eq:b_fit}
\end{align}
Setting $J/k_B = 52.0\,\text{K}$ and $T = 300\,\text{K}$,
Eq.~\eqref{eq:b_fit} provides an estimate of the magnitude of the
anisotropy, $|\delta| \approx 0.015$. We would like to point out,
however, that the error of $b$ is larger than $b$ itself, implying
that the estimate of $|\delta|$ from Eq.~\eqref{eq:b_fit} is not
reliable. This is mainly due to the fact that the anisotropy
$\delta$ of CPB is small and that the $y$ axis intercept $b$ is
proportional to $\delta^2$. However, at least an upper bound of the
order of magnitude can be estimated and agrees well with previous
findings of $\delta$. For other materials with larger anisotropy
this method would provide a way to estimate $\delta$ with a smaller
relative error.

We still use this value of $\delta$ to estimate the $g$-factor $g_c$
in $c$ direction from Eq.~\eqref{eq:m_fit}, since previously obtained
more reliable values, e.g.~$\delta = -0.019$, are close enough to
$\delta = -0.015$. Within the fit error of $m$, which is less
than $0.1\,\%$, these more reliable values would result in the same
value of $g_c = 2.153$. This $g$-value, in turn, is in excellent
agreement with $g_c=2.154$ obtained in the previous section by
fitting to angular-dependent data.

\subsubsection{\label{sec:HF-ESR_LT} Low temperatures}
\begin{figure}[t]
	\includegraphics[width=\columnwidth]{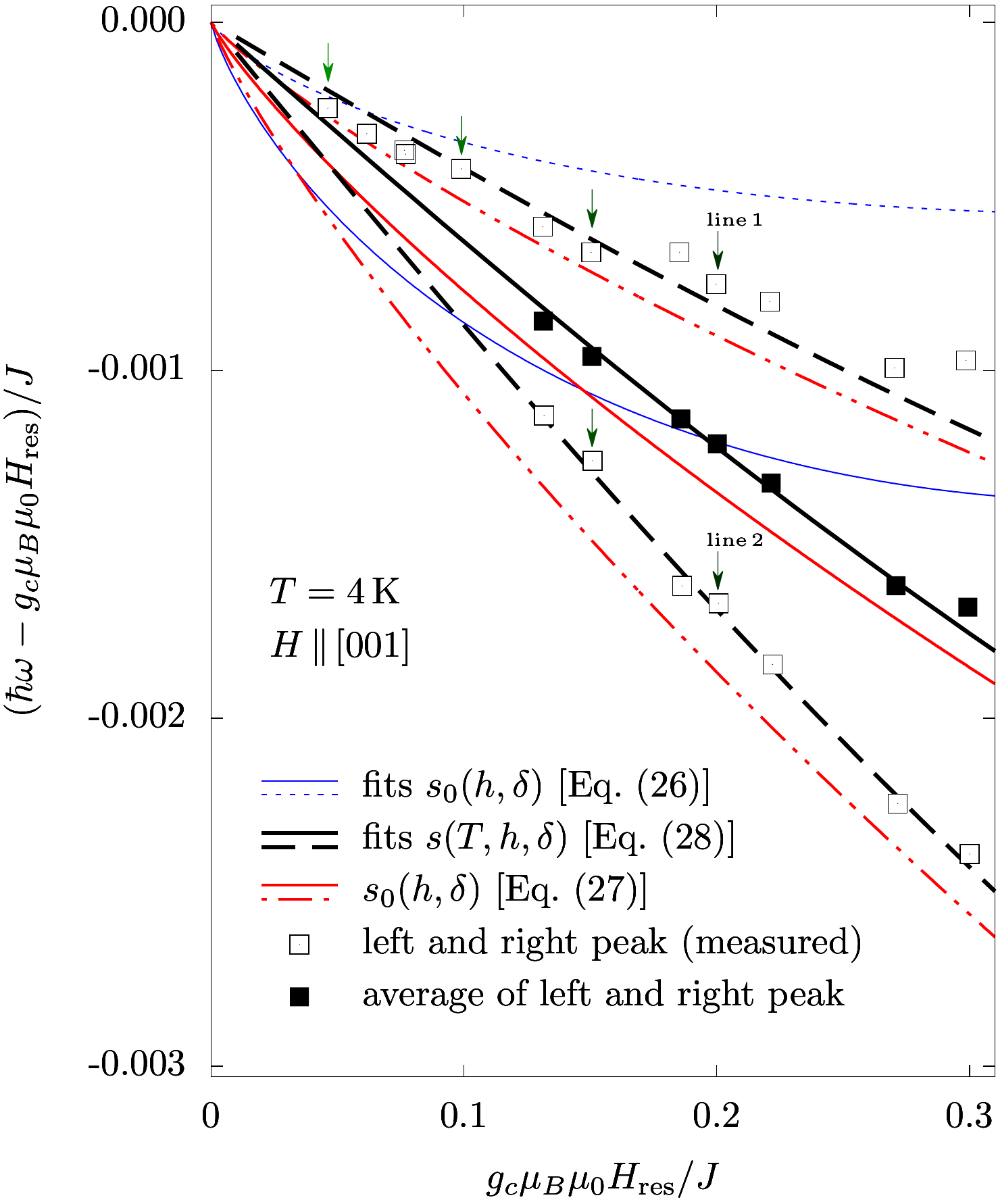}
	\caption{Resonance shift for CPB at $4\,\text{K}$ with magnetic
		field oriented along the $c$ axis, calculated from resonance 
		positions presented in Fig.~\ref{fig:CPB_HF-ESR_exp}. 
		Experimental data (squares) are compared with different 
		theoretical predictions (solid lines for the averaged shift, 
		dashed lines for the shifts of line 1 and line 2). Frequencies 
		and resonance fields are rendered dimensionless by multiplying 
		with $\hbar/J$ and $g_c\mu_B\mu_0/J$ (with temperature independent
		$g_c = 2.153$; see Sec.~\ref{sec:HF-ESR_HT}). Arrows 
		indicate the shifts of the exemplary spectra shown in the inset of 
		Fig.~\ref{fig:CPB_HF-ESR_exp}.}
	\label{fig:CPB_HF_ESR}
\end{figure}

The temperature-independent value of $g_c=2.153$ can be used in
the analysis of the resonance shift at low temperatures. Spectra
recorded at $4\,\text{K}$ consist of a resonance line (line 1 in
the inset of Fig.~\ref{fig:CPB_HF-ESR_exp}) present at all
frequencies and a second line (line 2) which evolves for higher
frequencies and is clearly visible above $141\,\text{GHz}$. Both
lines show an almost linear frequency-field dependence. The
deviations from a straight line can be attributed to the resonance 
shift, which is shown in Fig.~\ref{fig:CPB_HF_ESR}.

We compare three theoretical predictions for the resonance shift at
low temperatures, $T\ll J/k_B$, with the experimentally observed
data at $T=4\,\text{K}$ (black squares in Fig.~\ref{fig:CPB_HF_ESR}).
To this end, we subtract the dimensionless resonance fields
$h/J = g_c\mu_B\mu_0 H_\text{res}/J$ from the corresponding
dimensionless frequencies $2\pi\hbar\nu/J = \hbar\omega/J$. The result
defines the dimensionless ESR resonance shift $s(T,h,\delta)$.

The first prediction for the shift $s(T,h,\delta)$ was obtained 
by \mbox{Oshikawa} and Affleck within a field theoretical
approach\cite{Oshikawa2002} (blue lines in Fig.~\ref{fig:CPB_HF_ESR}).
It is supposed to hold for $T \rightarrow 0$ and reads
\begin{equation}\label{eq:s0_Oshikawa}
  s_0(h,\delta) := \lim_{T\to 0} s(T,h,\delta) \simeq \frac{h\delta }{J\pi^2}\ln\left(\frac{J}{h}\right)\,.
\end{equation}
The second prediction (red lines in Fig.~\ref{fig:CPB_HF_ESR})
is due to Maeda, Sakai, and Oshikawa.\cite{MSO05}
It extends Eq.~\eqref{eq:s0_Oshikawa} to larger resonance
fields as it includes logarithmic corrections to field theory,
\begin{equation}\label{eq:s0_Maeda}
  s_0(h,\delta) = \frac{h\delta}{J\pi^2}\left\{\mathcal{L} + \frac{\ln(\mathcal{L})}{2}  + \frac{3}{2} + \frac{1+\ln(\mathcal{L})}{4\mathcal{L}}\right\}
\end{equation}
with $\mathcal{L} = \ln[2J\sqrt{\pi^3}/(h\sqrt{2e})]$. This
equation was derived from the finite temperature result of
Ref.~\onlinecite{MSO05} (extended to arbitrary anisotropy
in Ref.~\onlinecite{Brockmann2011}) by taking the zero temperature
limit and expanding for small Zeeman energies $h$. In this work
we use a different definition of the resonance shift
(see App.~\ref{app:generalized_moments}) as compared to
Refs.~\onlinecite{MSO05} and \onlinecite{Brockmann2011}.
Up to first order in $\delta$, however, the resonance shift
at finite temperature is determined by the same combination
of static correlation functions,
\begin{equation}\label{eq:sT}
  s(T,h,\delta) = \delta\frac{\left\langle s_1^z s_2^z - s_1^x s_2^x \right\rangle_{T,h,0}}
                             {\left\langle s_1^z \right\rangle_{T,h,0}}\,.
\end{equation}
Finite temperature correlation functions as
$\left\langle s_1^z s_2^z\right\rangle_{T,h,0}$, $\left\langle s_1^x s_2^x\right\rangle_{T,h,0}$,
or the magnetization $\left\langle s_1^z \right\rangle_{T,h,0}$ per lattice
site of the isotropic spin-1/2 chain can be efficiently computed
using the quantum transfer matrix approach of Ref.~\onlinecite{Kluemper1992}
which reduces the problem to solving a finite set of well-behaved
non-linear integral equations (black lines in Fig.~\ref{fig:CPB_HF_ESR}).
Note that in Eqs.~\eqref{eq:s0_Oshikawa}, \eqref{eq:s0_Maeda}, and
\eqref{eq:sT} there is, in general, an angular dependent prefactor
$3\cos^2\vartheta - 1$ (see Eq.~\eqref{eq:m1_tilde_ang_dep} in
App.~\ref{app:ESR_parameters}) which is $-1$ here, since the
magnetic field is perpendicular to the anisotropy axis.

The three different theoretical curves \eqref{eq:s0_Oshikawa},
\eqref{eq:s0_Maeda}, and \eqref{eq:sT}, for several values of
$\delta$, are shown in Fig.~\ref{fig:CPB_HF_ESR}, where the
resonance shift is plotted as a function of $h/(g_c J) =
\mu_B\mu_0 H_\text{res}/J$ and compared with the experimental data.
We observe that Eq.~\eqref{eq:s0_Oshikawa} (solid and dashed
blue lines in Fig.~\ref{fig:CPB_HF_ESR}) is not fully consistent
with our experimental data. The best fit to the averaged shift
extracted from the two lines (full black squares in
Fig.~\ref{fig:CPB_HF_ESR}) over the full range of applied
resonance fields yields $\delta = -0.037$. On the other hand,
an extra\-polation of the experimental data for the shift
of line 1 (upper curve of open black squares) to small values
of $h/J$ and an asymptotic fit by eye of Eq.~\eqref{eq:s0_Oshikawa}
(dashed blue line) gives $\delta = -0.015$, which is compatible 
with our previous values. Eq.~\eqref{eq:s0_Oshikawa}
fails to explain the experimental data at higher resonance
fields because the validity of this formula is restricted
to $k_B T/J \ll h/J \ll 1$. But for the experimentally measured
resonance fields $\mu_0 H_{\rm res} \gtrsim 4\,\text{T}$,
i.e.~$h/(g_c J) \gtrsim 0.05$ (see Fig.~\ref{fig:CPB_HF-ESR_exp}),
the condition $h/J\ll 1$ is not sufficiently fulfilled.

In dimensionless units the temperature of $4\,\text{K}$ at
which our data were recorded translates to $k_B T/J \approx 0.08$.
Using Eq.~\eqref{eq:sT}, which is supposed to account of the
full temperature dependence and which is valid for all resonance
fields, the quality of the fit increases considerably (see
solid and dashed black lines in Fig.~\ref{fig:CPB_HF_ESR}).
The only free para\-meter in this case is the overall prefactor
$\delta$ in Eq.~\eqref{eq:sT}. A fit to the averaged shift
(full black squares), to the shift of line 1 (open black
squares, upper curve), and to the shift of line 2 (open black
squares, lower curve) implies $\delta = -0.012$, $\delta=-0.008$,
and $\delta=-0.017$, respectively.

For comparison, we also show Eq.~\eqref{eq:s0_Maeda} in
Fig.~\ref{fig:CPB_HF_ESR}. It includes higher corrections in
the magnitude of the resonance field but no temperature
corrections. The difference between Eqs.~\eqref{eq:s0_Maeda}
and \eqref{eq:sT} is therefore mostly due to the temperature.
In order to illustrate its effect we use the $\delta$ values
obtained from the fit of $s(T,h,\delta)$ in both cases.

In summary, we can infer from Fig.~\ref{fig:CPB_HF_ESR} that
the field theoretical result \eqref{eq:s0_Oshikawa} is insufficient
to explain our data for the field dependence of the resonance shift
in the full range $h/J \lesssim 0.3$. At least the
logarithmic corrections of Eq.~\eqref{eq:s0_Oshikawa} have to be
taken into account. The effect of small finite temperatures
($T = 4\,\text{K} \approx 0.077 J/k_B$) is clearly visible, and our
experimental data are better fitted and provide better (slightly
bigger) fit values of $\delta$ if the temperature dependence is
incorporated.

\subsection{\label{sec:temperature_dependence} Temperature dependence of ESR parameters}

In addition to the angular dependence of the ESR parameters at
room temperature and to the frequency dependence of the resonance
field at high and low temperatures we measured the temperature dependence
of the ESR parameters for $\boldsymbol{H}\,||\,[001]$ in two
different set-ups, first in the range between $4\,\text{K}$ and
room temperature at $9.56\,\text{GHz}$, and second for
temperatures between $1.6\,\text{K}$ and $300\,\text{K}$ at
$79.59\,\text{GHz}$.

\begin{figure}[t]
  \includegraphics[width=\columnwidth]{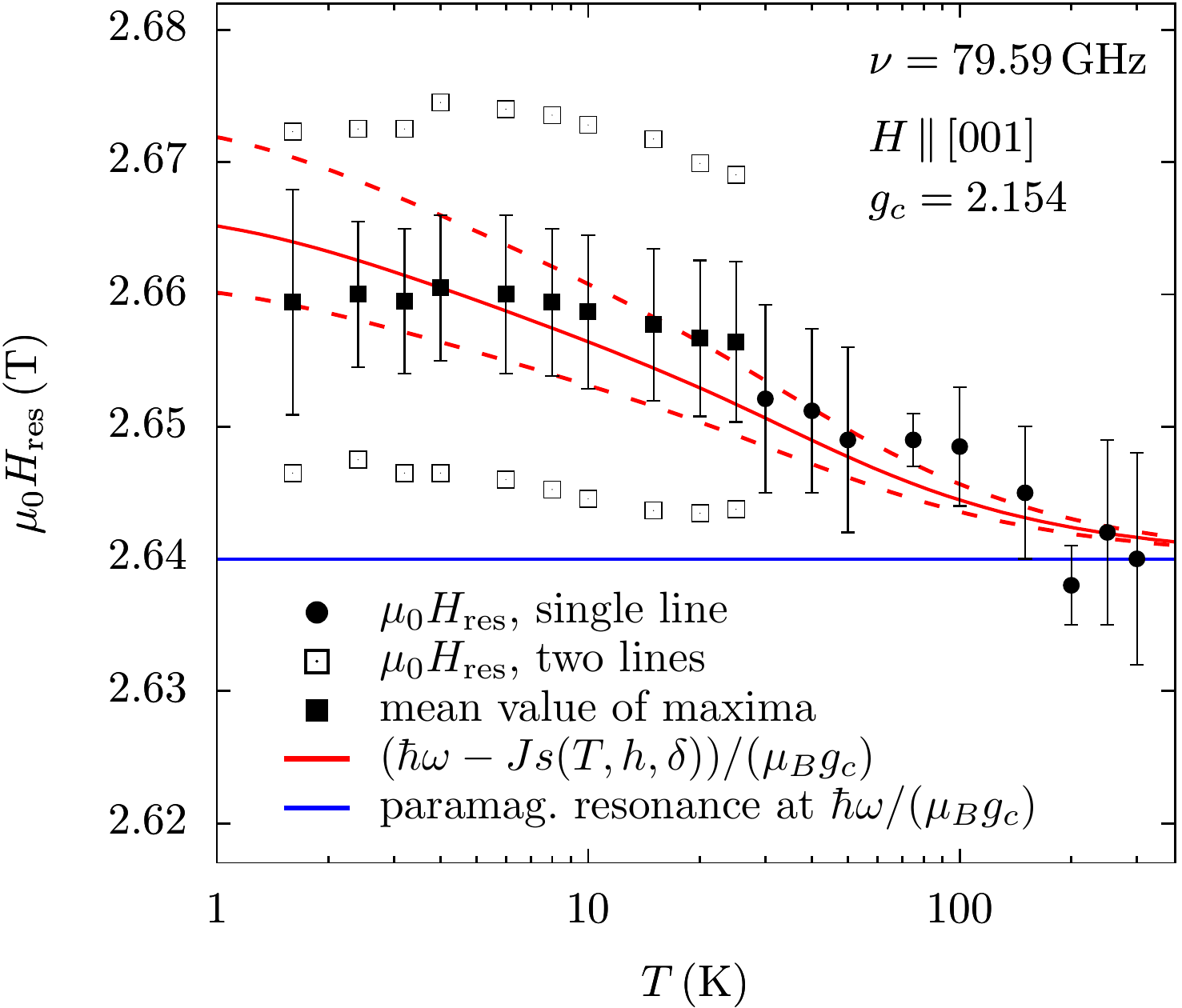}
  \caption{Resonance position for CPB as a function of temperature at $79.59\,\text{GHz}$ with the external magnetic
  field applied along the $c$ axis, compared with theoretical curves
  based on Eq.~\eqref{eq:sT}. The red solid line corresponds to $\delta=-0.015$,
  the dashed lines to $\delta=-0.012$ and $\delta=-0.019$.}
  \label{fig:Hres_CPB}
\end{figure}

The low-frequency measurements revealed only a very weak temperature
dependence of the resonance shift. From the HF-ESR measurements we
were able to extract the resonance shift with sufficient resolution
such that we could compare with Eq.~\eqref{eq:sT}. The result
is shown in Fig.~\ref{fig:Hres_CPB}. We find the agreement of the
theoretical prediction with our measured data quite remarkable
as no fitting was applied, and the values of the model parameters
$J$, $\delta$, and $g_c$ were taken from our previous measurements.
Note, in particular, that the correct sign of $\delta$ and the
proper angular dependence (factor $3\cos^2\vartheta-1$ in front of
the angular independent part of the resonance shift with
$\vartheta = 90^\circ$) are crucial in order to match experimental
and theoretical curves.

\begin{figure}[t]
  \includegraphics[width=\columnwidth]{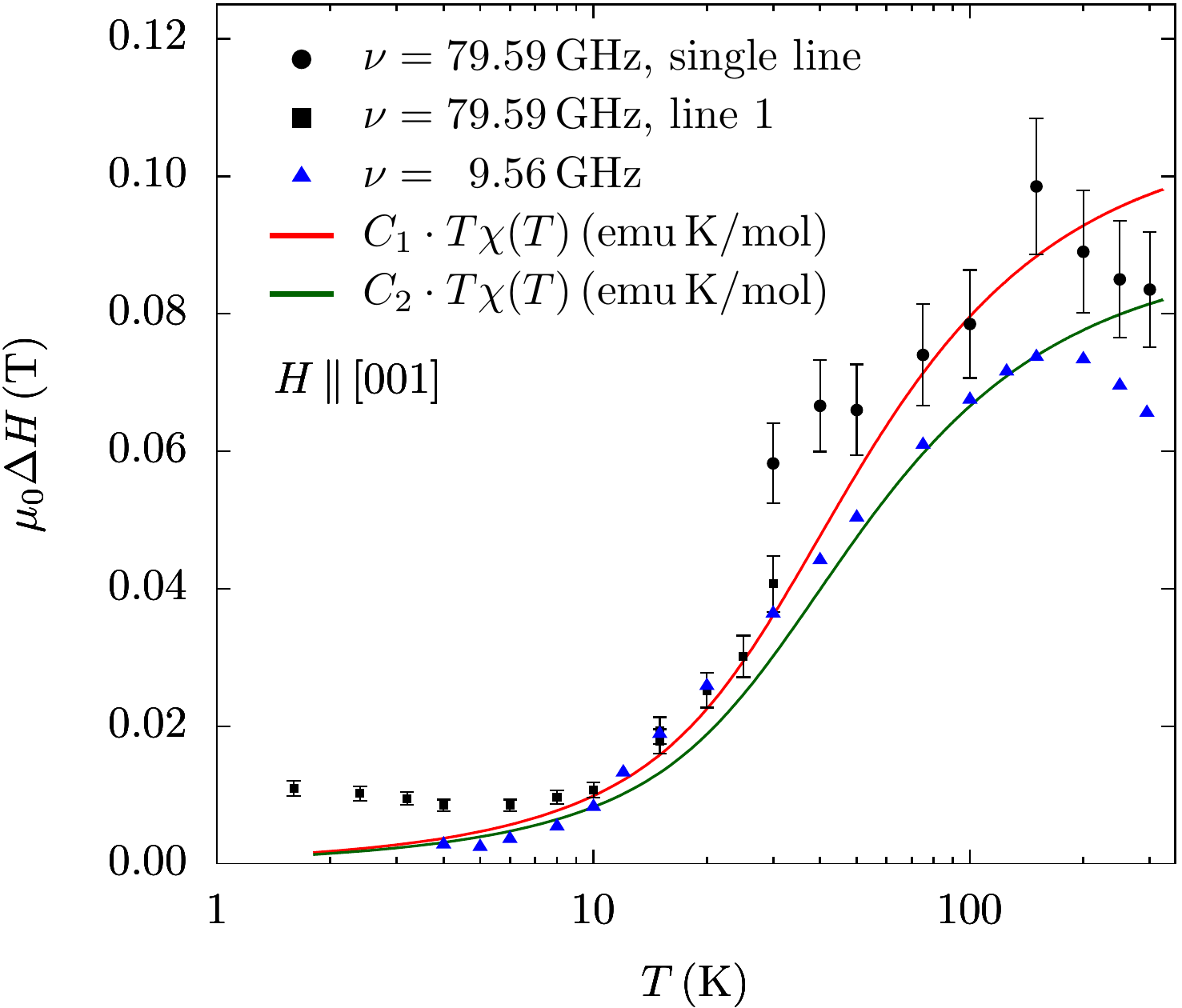}
  \caption{Linewidth for CPB as a function of temperature 
  at $9.56\,\text{GHz}$ and $79.59\,\text{GHz}$
  with the field applied along the $c$ axis. For
  comparison, the curves $C_{1,2} T\chi(T)$ with the 
  static susceptibility $\chi$ of the isotropic Heisenberg chain 
  and with $C_{1} = 0.25\,\text{T\,mol/(emu\,K)}$, 
  $C_{2} = 0.2\,\text{T\,mol/(emu\,K)}$ are shown.}
  \label{fig:CPB_Comparison_dH-ChiT}
\end{figure}

The linewidth as a function of temperature, as obtained in
low-frequency ESR, is shown in Fig.~\ref{fig:CPB_Comparison_dH-ChiT}.
Coming from high temperatures it increases until it reaches a
maximum of $74\,\text{mT}$ at around $150\,\text{K}$ and then
decreases rapidly with decreasing temperature. Below $10\,\text{K}$
this decrease is less steep and the linewidth reaches an apparently
constant value of $2.5\,\text{mT}$ at $4\,\text{K}$. The behavior
of the linewidth in our high-frequency experiment is very similar
for high and intermediate temperatures and is also shown in 
Fig.~\ref{fig:CPB_Comparison_dH-ChiT}.

For the temperature dependence of the linewidth we have no
reliable theoretical prediction so far. This is due to the parameter
values that characterize our compound, specifically due to the
very small value of the parameter $\delta$ which causes narrow
lines and would require a frequency resolution beyond the current
possibilities of our numerical method (see Ref.~\onlinecite{Brockmann2012}).
Analytical results for the full moments, on the other hand, are
available but can be only applied if the magnetic field is directed
along the anisotropy axis, which is impossible as we are dealing
with two inequivalent chains with anisotropy axes almost
perpendicular to each other (see Sec.~\ref{sec:angular_dependence}).
Moreover, these results do not compare well with the width at half
height as we have explained in Sec.~\ref{sec:general_remarks}
and in App.~\ref{app:ESR_parameters}.

In the framework of the phenomenological spin diffusion theory the
dynamics of the spin system is described by a diffusion equation.
For a one-dimensional system the linewidth is then expected to be
proportional to $T \chi(T)$ due to dominating $\boldsymbol{q} = 0$
fluctuations.\cite{Ajiro1975} The product $T \chi(T)$ is also shown
in Fig.~\ref{fig:CPB_Comparison_dH-ChiT}. We fitted the curves to
the data in the intermediate temperature range by adapting the
constant $C$ in $C T \chi(T)$. For $T \leq 150\,\text{K}$
the linewidth follows the $T\chi(T)$ behavior but considerably
deviates from it for temperatures above $150\,\text{K}$. These
findings hold for the low- as well as for the high-frequency
measurements. For the interpretation we should recall that spin
diffusion theory is a classical phenomenology which is expected to
give its best results in the high-temperature regime.

%

\section{\label{sec:neutron_scattering} Neutron scattering}

\begin{figure}[t]
	\includegraphics[width=\columnwidth]{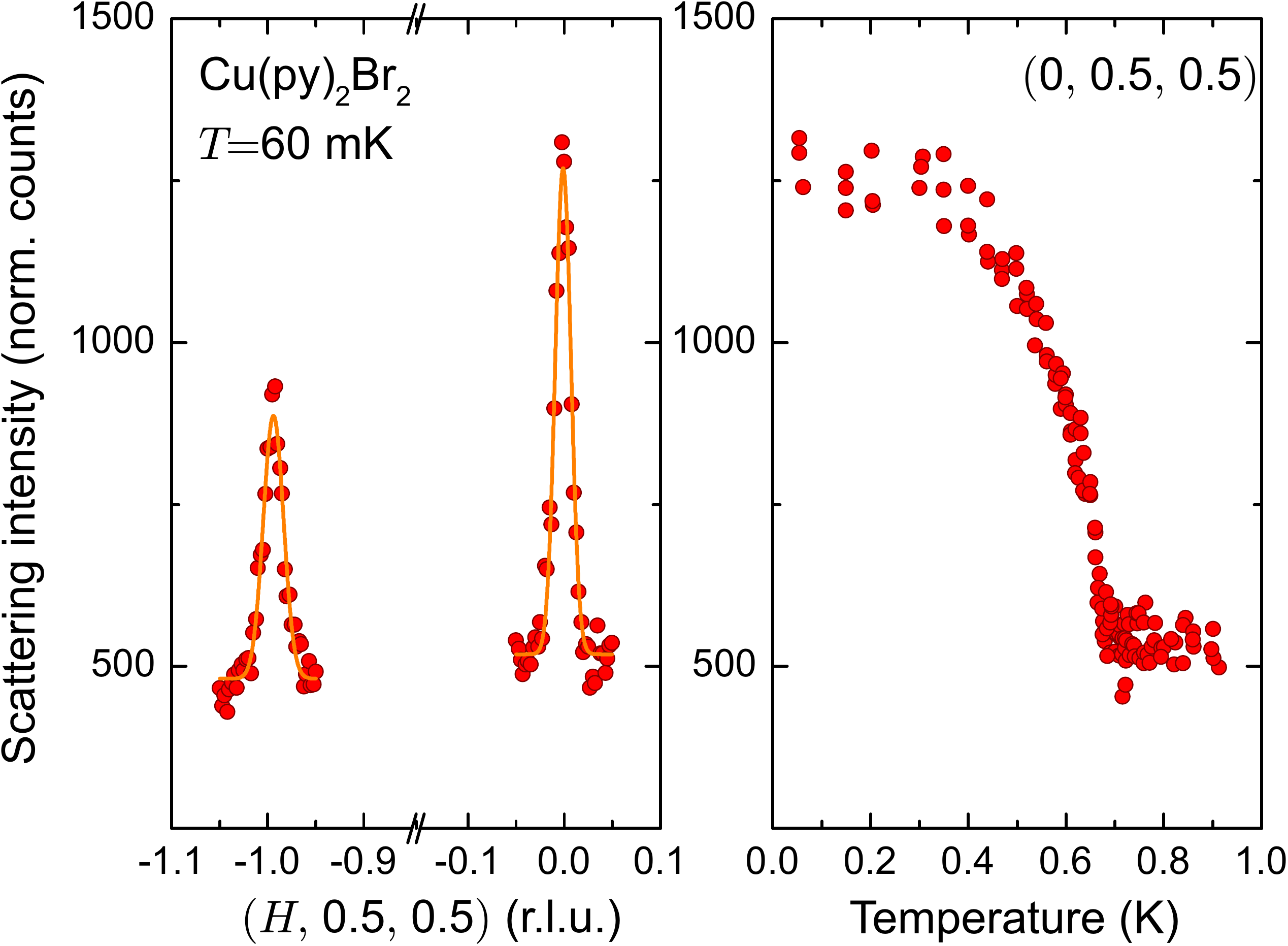}
	\caption{The results of neutron diffraction measurements on D23 instrument. Left: Neutron diffraction intensity along the ($H$,~0.5,~0.5) direction of the reciprocal space at the base temperature. Magnetic Bragg peaks of (0, 0.5, 0.5)-type
		are well pronounced. The solid lines are guide to the eye. Right: Diffraction intensity as function of temperature for (0,~0.5,~0.5) magnetic Bragg peak.
		Onset of magnetic scattering is visible around $T_{N}\simeq0.72$~K.}
	\label{fig:neutrondata}
\end{figure}

At $T_{N} \simeq0.72$~K the magnetic moments that are assigned to
the electron spins of the Cu$^{2+}$ ions in CPB order 
three-dimensionally. The low-temperature neutron diffraction 
experiment, after refining the lattice parameters of CPB to be 
approximately $a = 8.33$\,\AA, $b = 17.51$\,\AA, $c = 3.93$\,\AA, and 
$\beta = 96.6^{\circ}$ at $T=1\,\text{K}$, allowed us to establish 
the propagation vector of the magnetic structure $\mathbf{Q}=(0,~0.5,~0.5)$,
which implies a collinear ordering. Some corresponding magnetic 
Bragg peaks are shown in Fig.~\ref{fig:neutrondata}. They disappear 
around the same $T_{N}$ as the $\mu$SR and specific heat 
measurements suggest.\cite{Thede2012}

The observed propagation vector is fully consistent with the dominance
of antiferromagnetic intrachain interaction, $J>0$. Magnetic moments 
of nearest neighbors in $c$ direction prefer to align in an opposite
fashion. The body-centered arrangement of spins within the unit cell
leads to a perfect frustration between the two chain subtypes.
Probably, it could be resolved via taking the quantum fluctuations
into account. Such order-by-disorder type of mechanism is known to
select the most collinear arrangement from the degenerate manifold
of states.\cite{SizanovSyromyatnikov_JPCM_2011_coupledsublattices}
A similar example of system with interpenetrating collinear magnetic
sublattices and perfect frustration between them is found in the 
$S=1$ quantum magnet DTN.\cite{SizanovSyromyatnikov_JPCM_2011_coupledsublattices, Tsyrulin_JPCM_2013_DTNneutr}
We thus propose a fully collinear arrangement of spins in CPB. 
The tentative structure is shown in Fig.~\ref{fig:CPB_structure}.

This low-temperature spin structure is supported by our analysis of 
ESR data. Due to negativity of $\delta$ the spins prefer to align 
inside the plane perpendicular to the anisotropy axis (see
Sec.~\ref{sec:angular_dependence}) which is for each chain the plane
defined by the bromine ions (see Fig.~\ref{fig:CPB_structure}). For
two of the inequivalent chains these planes are almost perpendicular
to each other. This leaves the antiferromagnetic arrangement of the
spins along the chain as the only plausible choice for the
structure (see App.~\ref{app:spin_structure}). 

%

\section{\label{sec:doping} \texorpdfstring{Cu(py)$_2$(Cl$_{1-x}$Br$_x$)$_2$}{CPC->B}: Impact of doping}

Samples with two different Cl concentrations of $2\%$ ($x = 0.98$)
and $5\%$ ($x = 0.95$) were investigated in order to study the
influence of doping at the halogen sites on spin dynamics. For both
systems the angular dependence of the ESR parameters at room
temperature as well as their temperature dependence was measured at
$9.56\,\text{GHz}$. The former measurements are qualitatively
similar to the results obtained for CPB and are not discussed any
further. As in the case of the pure compound, the angular
dependence could be used to identify the crystallographic $c$ axis.
Measurements of the temperature dependence were performed with
magnetic field applied along the $c$ axis in the range between
$4\,\text{K}$ and room temperature. \mbox{A small} shift of the resonance
fields to higher values with decreasing temperature was observed,
similar to CPB.

\begin{figure}[t]
	\includegraphics[width=\columnwidth]{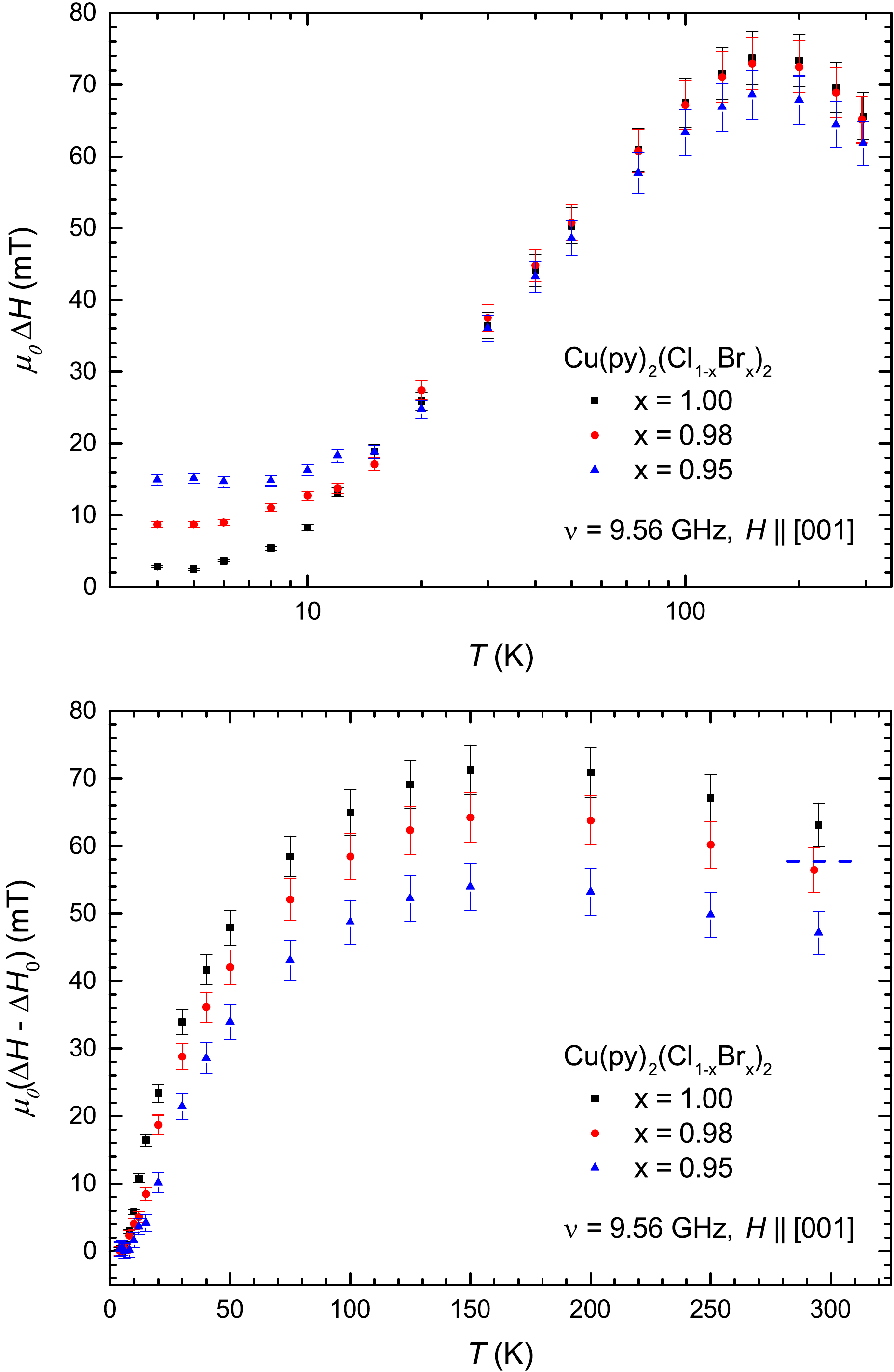}
	\caption{Temperature dependence of the ESR linewidth measured at
		$9.56\,\text{GHz}$ for $x = 1.0, 0.98, 0.95$ (top) and of the
		contribution governed by spin dynamics obtained after subtracting
		the contribution of the inhomogeneous broadening (bottom).
		Note the difference in scales used in the upper and lower panels. 
		The magnetic field was applied along the $c$ axis. The blue 
		dashed line indicates the linewidth for $x=0.95$ as expected 
		from a relative change in $J$ only.}
	\label{fig:CPX_X-Band_T-dep}
\end{figure}

The linewidth as a function of temperature is shown in the upper
panel of Fig.~\ref{fig:CPX_X-Band_T-dep} for all three systems
studied in this work. Qualitatively, the behavior of the linewidth
is the same for the three compounds. However, a constant
low-temperature linewidth increases with increasing Cl
concentration while at high temperatures this trend is reversed,
i.e.~the undoped compound shows the largest linewidth. A possible
reason for this behavior lies in the different contributions to the
linewidth. 

The disorder in the crystals increases with increasing Cl content. 
Thus, the inhomogeneous broadening of the resonance lines, most 
likely caused by the local and spatially varying alteration 
of the $g$-tensor, increases with doping concentration as 
well. This effect is temperature independent and dominates the 
linewidth at low temperatures, thereby explaining the observed 
changes of the low-temperature linewidth.

The second contribution is given by spin dynamics of the system
whose temperature dependence can be studied by subtracting the
contribution of inhomogeneous broadening $\Delta H_0$ from the
data. In the lower panel of Fig.~\ref{fig:CPX_X-Band_T-dep}
linewidths are shown after subtraction. Note the use of a linear
temperature scale in this graph which is better suited for the
following discussion. In the high-temperature regime,
Eq.~\eqref{eq:HTE_linewidth} holds and describes a linear relation
between linewidth and isotropic exchange. In Ref.~\onlinecite{Thede2012}
the strength of effective isotropic exchange was determined for
CPB, CPC and various mixed compounds with different Cl and Br
contents. It was found that isotropic exchange monotonically
decreases with increasing Cl doping and is minimal for CPC. This is
in qualitative agreement with the observed decrease of linewidth
for increasing Cl content. Quantitatively, however, the relative
change in $\Delta H$ is larger than the relative change in $J$.
This is illustrated by a dashed horizontal line in the lower part
of Fig.~\ref{fig:CPX_X-Band_T-dep} which indicates the linewidth
of the $5\,\%$ doped sample at $300\,\text{K}$ as expected from
the change in $J$. Thus, the behavior of the linewidth cannot
solely be described in terms of change in isotropic exchange. 

A possible explanation of this finding could be an effective
decoupling of anisotropic exchange from isotropic exchange, meaning
that in Eq.~\eqref{eq:HTE_linewidth} $J\delta$ might vary independently 
of $J$ as functions of doping. The existence of such a decoupling was 
shown theoretically\cite{Tornow1999} and experimentally\cite{Kataev2001} 
in the case of chains with ferromagnetic exchange coupling.
Note that the Cu-Br-Cu bond angle of the superexchange path is
$89.64^{\circ}$, i.e.~close to $90^{\circ}$ for which one would
expect a ferromagnetic exchange.\cite{Goodenough1955,Kanamori1959,Anderson1963}
The small deviation from $90^{\circ}$ leads to an antiferromagnetic
but relatively weak isotropic exchange in accordance with the
experimentally determined value.

\section{\label{sec:discussion} Discussion}
In previous sections we presented a combined experimental and 
theoretical study of the magnetic properties of the spin chain 
compounds Cu(py)$_2$(Cl$_{1-x}$Br$_x$)$_2$ ($x$ = 1.0, 0.98, 0.95). 
We begin the discussion of the obtained results by considering ESR 
measurements of CPB performed at low frequencies and room 
temperature. From studies of the angular dependence of resonance 
position and linewidth we inferred the existence of two distinct 
anisotropy axes in this system. These axes are related to the two 
magnetically inequivalent chain types and are oriented almost 
perpendicular to each other within the plane perpendicular to the 
chain axis. Moreover, they coincide with the axis formed by the two 
nitrogen ligands of the respective local octahedral environment of 
the Cu$^{2+}$ ions (see Fig.~\ref{fig:coordinate_systems}). The 
insight into number and orientation of anisotropy axes in this 
material is an important finding, as it is an essential ingredient 
for modeling of our data. 

Combining this knowledge with novel expressions for the angular 
dependence of resonance field and linewidth of individual chains 
(Eqs. \eqref{eq:HTE_shift} and \eqref{eq:HTE_linewidth}, 
respectively) we were able to describe the observed angular 
dependence of both ESR parameters (see Figs.~\ref{fig:X-band_ang_dep_I} 
and \ref{fig:X-band_ang_dep_II}). Thereby, we could determine the 
complete $g$-tensor of the pure compound CPB. The $g$-factors 
obtained from the fit to our measured data of the angular 
dependence of resonance field and linewidth, Eq.~\eqref{eq:g_fit}, 
agree well with the values reported in Ref.~\onlinecite{Pal1994} 
for measurements of the $g$-factor angular dependence in the $a$-$c$ 
and $b$-$c$ planes. In Ref.~\onlinecite{Pal1994} Pal et~al.\ found 
for the $g$-factor in $b$ direction $g_b=2.065$ which coincides 
with our value. The minimal and maximal values of the (effective) 
$g$-factor when rotating the field in the $a$-$c$ plane were found 
to be $g_\text{ac}^{\rm (max)} = 2.178$ and $g_\text{ac}^{\rm (min)} = 2.056$
around angles of $-20^\circ$ and $70^\circ$, respectively (labeled
by $g_1$, $g_2$, and $\Psi$ in Tab.~1 of Ref.~\onlinecite{Pal1994}).
By means of our fully determined $g$-tensor we obtain
$g_\text{ac}^{\rm (max)} = 2.175$ and $g_\text{ac}^{\rm (min)} = 2.047$
around $-25^\circ$ and $+65^\circ$. The differences in $g$-factors
might be attributed to two facts. First, the authors of Ref.~\onlinecite{Pal1994}
do not take into account the resonance shift due to the anisotropic 
exchange of the many-body system. Secondly, they assume a $g$-tensor 
of cylindric symmetry (only $g_\|$ and $g_\perp$ in Ref.~\onlinecite{Pal1994}) 
whereas we consider it to be more general with three different 
eigenvalues $g_1$, $g_2$, and $g_3$. On the other hand, we assume 
the principal axes of the $g$-tensor to coincide with the symmetry 
axes of the local octahedral environment of the Cu ions, whereas in
Ref.~\onlinecite{Pal1994} the angle of the maximum position of 
$g_\text{ac}$ is fitted and disagrees by about $5^\circ$ from our 
angle. Considering the width of the observed resonance lines, their 
data of `peak-to-peak' linewidths (see Fig.~4 in Ref.~\onlinecite{Pal1994}) 
agree well with our data for widths at half height shown in Fig.~\ref{fig:X-band_ang_dep_II}. 
The difference is a factor of about $\sqrt{3}$ which is typical for
Lorentzian-like lineshapes. Thus, our results are fully consistent 
with previously published studies.

Furthermore, by fitting the angular dependence of the linewidth,
we could derive the exponent $\gamma \approx 0.7$ of the algebraic
long-time decay of a certain correlation function of the
isotropic model at room temperature ($T\approx 6J/k_B$). This correlation
function describes the propagation of two neighboring spin
flips through the isotropic chain and enters our theory through
the perturbation expansion in the anisotropy parameter $\delta$.
It is worthwhile mentioning, that an analysis based on
Eq.~\eqref{eq:HTE_linewidth} is by no means restricted to the
specific system which is discussed here. Thus, our findings may
serve for the investigation of other close-to-isotropic 1d systems
thereby giving insight into their spin dynamics. Finally, the 
fits to our angular dependent data yielded a range of reasonable 
values for the anisotropy parameter,  $-0.04 \leq \delta \leq -0.01$.

This information on $\delta$ was confirmed, and even more 
specified, by a detailed analysis of magnetization measurements 
which were performed on a CPB crystal for external fields applied 
parallel and perpendicular to the spin chains. The model used
for the analysis takes into account the anisotropy $\delta$ of the 
system as well as the specific orientation of the two anisotropy 
axes. Therefore, it extends the existing descriptions of isotropic
1d chains like, for instance, the one employed in the approach
of Ref.~\onlinecite{Johnston2000}. Compared to values reported
in literature, we obtained a refined value of the intrachain
coupling strength $J=4.48\,\text{meV}$, i.e.~$J/k_B = 52.0\,\text{K}$,
as well as the anisotropic exchange coupling $J\delta \approx -0.09\,
\text{meV}$, i.e.~$\delta\approx -0.02$, which was unknown up to 
now. The value of $J$ is close to the previously reported 
value\cite{Thede2012} of $J \approx 4.58\,\text{meV}$. In any case, 
it improves estimates obtained in Refs.~\onlinecite{VanOoijen1977} 
and \onlinecite{Jeter1972}, where the authors found $J/(2\pi\hbar c) = 33.2\,\text{cm}^{-1}$ 
and $J/(2\pi\hbar c) = 37.8\,\text{cm}^{-1}$, respectively,
i.e.~$J/k_B = 48\,\text{K}$ and $J/k_B = 54\,\text{K}$, both with
errors of the order of $5\%$.

We emphasize that our procedure of estimating the anisotropy
from two susceptibility measurements with different field
directions is not limited to the special compound CPB. The method
works for any close-to-isotropic model with a small anisotropic
perturbation $V$ for which the thermal expectation value of the
perturbation term can be computed. This is explained in detail
in App.~\ref{app:susceptibility_general}. In
App.~\ref{app:susceptibility_example} we also present a simplified
method to estimate the anisotropy as well as the isotropic
intrachain coupling strength which is only based on the ratio of
the temperatures at the maxima of the two susceptibility curves.
In the case of an isotropic system the well-known exact result of 
Ref.~\onlinecite{Johnston2000} is reproduced by the novel 
procedure. Applying the latter to our data measured on CPB, we 
obtained $\delta \approx -0.03$ and $J/k_B\approx 52.2\,\text{K}$, 
which is in a good agreement with values resulting from fitting 
magnetization data over almost the whole experimentally available 
temperature range. Thus, we provided expressions which might prove 
to be useful for an easy estimation of $J$ and $\delta$ in related 
systems with anisotropies being not too large.

Besides measurements of magnetization and ESR properties at low 
frequencies we performed a HF-ESR study on CPB in order to 
investigate the behavior of the resonance shift as a function of 
magnetic field and temperature in more detail. At high temperatures, 
recorded spectra consist of a single resonance line. From frequency 
dependent measurements at $300\,\text{K}$ we extracted the 
temperature-independent $g_c$-factor, $g_c=2.153$ (see 
Sec.~\ref{sec:HF-ESR_HT}), which is very close to the X-band result 
$g_c=2.154$ and which is typical for Cu$^{2+}$ ions in an octahedral 
environment.\cite{Abragam_Bleaney} Afterwards, this value for $g_c$ 
was used for calculating the resonance shift at low temperatures, 
as it is discussed below. Moreover, we aimed at extracting an 
additional independent value of the anisotropy parameter from 
these data. As it turned out, the method of determining the 
anisotropy parameter $\delta$ from data for the resonance position 
at high temperature as a function of frequency is not reliable in 
the case of CPB since its anisotropy is too small ($|\delta|\approx 0.02$). 
However, since the quantity from which $\delta$ is estimated is 
proportional to $\delta^2$, we believe that it provides better 
estimates if $|\delta|$ is bigger, e.g.~$|\delta|\gtrsim 0.1$. 
Therefore, the presented analysis may find further applications to 
systems beyond the scope of this work.

In contrast to high temperatures, low temperature HF-ESR spectra of 
the undoped sample contain two lines visible at high frequencies.
For the explanation of the appearance of these two spectral lines 
we favor a scenario based on the existence of a (small) intergrown 
crystal with slightly different orientation of its $c$ axis. This 
would explain the different spectral weights of the two peaks (with 
the intensity of the smaller peak proportional to the volume of the
intergrown crystal) as well as the different positions of the peaks
(corresponding to different $g$-factors due to different angles
between magnetic field and the two $c$ axes). This scenario is
further supported by the fact that we could not observe any
double-peak structure in our HF-ESR measurements on the doped
samples (data not shown) which in most other respects behave
qualitatively similar to the undoped sample (see Sec.~\ref{sec:doping}). 
Another possible scenario, which cannot be fully ruled out, is that
the two lines can be attributed to the two magnetically
inequivalent chains. However, within this scenario we also would
expect two lines of equal intensity which is in contrast to the
experimental findings, rendering this scenario less likely.
As it is not possible to determine the origin of the two lines 
definitely, we took into consideration the resonance positions of 
both lines as well as the mean resonance field for our 
investigation of the low temperature resonance shifts. 

The deviations from a straight line as found in our HF-ESR
measurements at $4\,\text{K}$ (see Fig.~\ref{fig:CPB_HF_ESR}) could
be explained by a low-temperature formula for the resonance shift,
yielding a negative value of the anisotropy as well as an estimate
of its magnitude $\delta = -0.012$. Furthermore, by comparing
formulae stemming from different approaches we could show that
the field theoretical result \eqref{eq:s0_Oshikawa} is insufficient
to explain our data for frequencies $\hbar\omega/J \gtrsim 0.1$.
This evidences the importance of logarithmic corrections
as in Eq.~\eqref{eq:s0_Maeda} for describing the resonance
shift in magnetic fields which do not fulfill the condition
$h/J \ll 1$ as it is the case in our study. An even better agreement
between experimentally obtained and calculated shifts was found, if
we take into account finite temperature effects, cf.~Eq.~\eqref{eq:sT},
which are visible in our data despite the fact that measurements
were performed for $T = 0.08 J/k_B$.

Our temperature-dependent ESR data for the resonance shift 
(see Fig.~\ref{fig:Hres_CPB}) agree very well with the theoretical 
prediction \eqref{eq:sT}, using the previously obtained values 
$J/k_B = 52.0\,\text{K}$, $g_c=2.154$, and $\delta\approx -0.02$
without any fitting. However, it seems that the data at very low 
temperatures ($T\leq 4\,\text{K}$) are better matched by assuming 
small anisotropies, e.g.~$|\delta|=0.012$, $0.015$, instead of 
$|\delta|\approx 0.02$. On the other hand, the overall temperature 
dependence, in particular at low and intermediate temperatures, 
$0.1 \leq k_B T/J \leq 3$, can be well explained by 
assuming a larger anisotropy, e.g.~$|\delta| = 0.019$ (as obtained 
from our susceptibility measurements). This fits with the fact
that our derivation of Eq.~\eqref{eq:sT} in App.~\ref{app:ESR_parameters}
assures its validity if the the condition $\delta \ll k_B T/J$ is
satisfied, while the extension to lower temperatures is based on
more hand-waving arguments.

Furthermore, by combining the information on $\delta$ as well
as on the existence of two distinct anisotropy axes with the
information about the propagation vector of the ordered state,
as obtained from neutron scattering experiments, a tentative
spin structure at zero temperature could be proposed (see
Fig.~\ref{fig:CPB_structure}). Strong theoretical support for
this structure was obtained from a renormalization group argument
(see App.~\ref{app:spin_structure}). Thus, the present 
study is an example for combining ESR measurements,
neutron scattering experiments, and theoretical arguments as
complementary methods to gain information about the
spin structure of the ordered state of a real physical system.

Considering the anisotropy parameter, we obtained at least three 
reliable and independent estimates for $\delta$ based
on magnetization and ESR measurements which allowed us to
establish the strength and the sign of the anisotropy of CPB
to be $\delta \approx -0.02$. The latter finding is an important
contribution to the evaluation of the utility of CPB as a
realization of the XXZ model. As was mentioned
in Sec.~\ref{sec:general_remarks}, the XXZ model belongs to
the class of integrable lattice models. This fact makes it
possible to calculate its thermodynamic properties and some
of its correlation functions exactly for the infinite chain.
Unfortunately, an external magnetic field generally breaks
the integrability, unless it is applied in the direction of
the anisotropy axis. Our finding that the anisotropy axes of
the two inequivalent chains in CPB are oriented perpendicular
to each other and perpendicular to the chain axes makes it
impossible to apply any of the known exact results for the
XXZ model, except when the external field is switched off, as any
finite field will necessarily be non-parallel to at least the
anisotropy axis of one of the two families of inequivalent
chains. For the applicability of the results obtained in Ref.~\onlinecite{Brockmann2011}, 
for instance, we would have needed that
the anisotropy axes would be oriented along the chain
direction. What may be seen as bad luck with the orientation
of the anisotropy axes was somewhat compensated by our finding
that the anisotropy parameter is very small in modulus. This
fact allowed us to perform a first order perturbation theory
in $\delta$ and to use exact results for the isotropic Heisenberg
chain which remains integrable for arbitrary direction of
the applied magnetic field. As the comparison with the
experimental results shows, this works very well for the
susceptibility of the XXZ chain and for the description of
the temperature dependence of the resonance shift. We have
further combined the perturbative expansion in $\delta$ with
a high-temperature analysis and with an analysis of the
cut-off dependence of modified moments which gave us access
to the angular dependence of the linewidth in the high-temperature
regime. Altogether the challenges provided by the experimental
data have inspired the develop\-ment of new ideas on the theory
side whose applicability is not restricted to our specific
example but can also be applied to take into account, for instance, 
small XYZ anisotropy, small next-to-nearest neighbor coupling,
or the coupling of adjacent chains.

In order to verify the prediction\cite{Brockmann2011} of a strong 
deviation from the linear dependence of the ESR resonance shift on 
the magnetic field in XXZ magnets, not only a material with a 
single anisotropy axis would be required, but we would need a 
material with smaller $J$ and larger $\delta$. In such a material 
we would also have a chance to reliably calculate cut-off dependent 
moments numerically which would give us direct access to the 
experimentally measured linewidth at half height.

\section{\label{sec:conclusions} Conclusions}

A detailed theoretical analysis of the experimental data presented 
in this paper has shown that the magnetic properties of CPB as seen 
in ESR and magnetization measurements can be well understood within 
the following simple picture tightly connected with the crystal 
structure of this compound. The copper ions in the crystal form 
antiferromagnetic spin-1/2 chains with an exchange coupling of $J = 4.48\,\text{meV}$.
The local environment of a magnetic ion consists of four bromine 
and two nitrogen ligands which form a stretched octahedron. As a 
consequence of the asymmetry of this local environment, the three 
eigenvalues of the $g$-tensor, whose principal axes coincide with 
the symmetry axes of the stretched octahedron, are mutually 
different, $(g_1,g_2,g_3) = (2.065,\,2.018,\,2.203)$. 

Furthermore, the isotropic exchange interaction is distorted by a 
small anisotropic component. This component is well accounted for 
by a small Ising interaction of strength $J \delta = -0.09\,\text{meV}$ 
directed perpendicular to the bromine planes. As there are two 
types of octahedra in the material, which map onto each other by a 
glide reflection, there are two inequivalent spin chains whose 
anisotropy axes are (almost) perpendicular to each other and 
perpendicular to the chain direction. As compared to the intrachain 
coupling $J$, the interchain interaction is weak as can be seen 
from the small value of the ordering temperature $T_N=0.72\,\text{K}$, 
confirmed by neutron scattering experiments. 
Applying renormalization group arguments to the model of two weakly
coupled XXZ chains (see App.~\ref{app:spin_structure}) we suggested
a magnetic structure in the ordered phase ($T<T_N$) that is consistent
with the propagation vector $\mathbf{Q}=(0,\,0.5,\,0.5)$ obtained
from neutron scattering experiments and consists of antiferromagnetically
ordered collinear spins oriented along the chain direction. From
the dependence of the linewidth on doping concentration we found
evidence for an effective decoupling of the anisotropic component
$J\delta$ from the isotropic exchange $J$ as function of doping.
Here the details remained open. Their explanation would require
further theoretical studies. 

On the theoretical side of this work, we have developed an approach
to estimate the exchange anisotropy $\delta$ from static magnetization
measurements with fields applied in different directions (see 
App.~\ref{app:susceptibility}), which is
applicable to spin chain compounds with small anisotropies. Our
analysis of the ESR data relied on a novel approach to computing
ESR parameters from moments of the dynamical susceptibility with an
inherent cut-off in frequency (see App.~\ref{app:ESR_parameters}).
Since this approach connects the angular dependence of the linewidth
with the algebraic decay in time of a certain correlation function
of the isotropic Heisenberg chain, we were able to determine the
corresponding exponent $\gamma$ for high temperatures experimentally.

For the future we hope from the experimental side for the 
development of more efficient theoretical methods for the 
computation of dynamical correlation functions at finite 
temperature which would allow us to obtain a better prediction for
the behavior of the experimental linewidth at all temperatures. Our 
hope from the theoretical side is that the search for experimental 
systems with simpler geometry (such that an alignment of the 
magnetic field along a single anisotropy axis is possible) will be 
successful. At the same time, spin chain materials with bigger 
anisotropy and smaller $J$ (such that higher effective fields $h/J$ 
are accessible) are much sought after. These are expected to show 
an interesting non-monotonic behavior\cite{Brockmann2011} 
of the resonance shift as a function of the external field.

\section*{Acknowledgements}

We would like to thank Jesko Sirker and Xenophon \mbox{Zotos} for valuable 
discussions. Furthermore, we would like to thank Christian Blum for 
technical support with orienting the CPB crystals. MB is grateful 
to the Max Planck Institute for the Physics of Complex Systems in 
Dresden where parts of this project were done. Theoretical parts of 
the work were carried out in the framework of the DFG research 
group FOR-2316. Experimental work at the IFW was partially supported 
by project KA1694/8-1 funded by the DFG. Neutron measurements were 
supported by Swiss National Science Foundation, Division~II.

\appendix

\section{\label{app:susceptibility} Static susceptibility of close-to-isotropic models}

In this section we proposse a new method to obtain quantitative
estimates of anisotropic perturbations of magnetically isotropic
many-body systems by means of magnetization or susceptibility
measurements. For simplicity we will consider magnetization and 
susceptibility as scalars. This is motivated by the fact that the
magnetic field direction is often chosen (almost) parallel to
one of the principal axes of the susceptibility tensor. The
static zero-field susceptibility $\chi(T)$ can then be expressed
by the magnetization $m(T,h)$ per lattice site,
\begin{equation}\label{eq:chi_def}
  \chi(T) = \partial_h m(T,h)\Big|_{h=0} \approx \left.\frac{m(T,h)}{h}\right|_\text{$h$ small}\,,
\end{equation}
where the Zeeman energy $h=g\mu_B\mu_0 H$ is proportional to the
strength $H$ of the magnetic field. Note that the first relation in 
\eqref{eq:chi_def} implies that $J\chi(T)$ is dimensionless, like
the magnetization $m(T,h)$ itself. Standard units as in Eqs.~\eqref{eq:chi1_corr} 
and \eqref{eq:chi2_corr} can be restored in the end. The second 
relation in \eqref{eq:chi_def} assumes a linear dependence of the 
magnetization on the applied field which is typically justified in 
antiferromagnets if the field is not too large. The magnetization
of CPB, discussed in the main body of the text, for instance,
was measured in small residual fields of about $0.1\,\text{T}$.

The main idea to be worked out below is to measure the static
zero-field susceptibility $\chi(T)$ (or equivalently the
magnetization for small fields) as a function of temperature
for different magnetic field directions. A comparison of
the susceptibility profiles then allows us to gain information
about the magnetic anisotropy of the perturbation.

\subsection{\label{app:susceptibility_general} Perturbation expansion of the magnetization}

We consider a Hamiltonian of the form
\begin{equation}\label{eq:Hamiltonian_perturb}
  \mathcal{H} = \mathcal{H}_0 + \mathcal{H}_Z + \lambda V
\end{equation}
with Zeeman term $\mathcal{H}_Z = - h \bm S \cdot \ev$ and $SU(2)$-symmetric
Hamiltonian $\mathcal{H}_0$. The operator $V$ characterizes the
anisotropic perturbation. We assume that both, $\mathcal{H}_0$ and $V$,
have the same typical energy scale $J$ and that $\lambda$ is a small
dimensionless number. Furthermore, $\ev = \hat{g}\bm H/|\hat{g}\bm H|$,
where $\hat{g}$ is the $g$-tensor, will denote a unit vector in field
direction, $\bm S = \sum_{j=1}^L \bm s_j$ is the total spin, and
$h = \mu_B\mu_0|\hat{g}\bm H|$ is the Zeeman energy corresponding to
the magnetic field $\hat{g} \bm H$. Then, the component in field
direction of the dimensionless magnetization per lattice site reads
\begin{align}
  m(T,h,\lambda) &= \frac{k_B T}{L} \partial_h \ln\left[\text{Tr}\left\{e^{-\mathcal{H}/(k_B T)}\right\}\right]\,. \label{eq:magnetization_def}
\end{align}
A perturbation expansion up to first order in $\lambda$ yields
\begin{align}
  m(T,h,\lambda) &\simeq m(T,h,0) - \frac{\lambda}{L}\partial_h \langle V \rangle_{T,h,0}\label{eq:magnetization_perturb}\,,\\
  \Rightarrow\quad \chi(T) &\simeq \chi^{(0)}(T) - \frac{\lambda}{L} \lim_{h\to 0}\partial_h^2 \langle V \rangle_{T,h,0} \label{eq:susceptibility_perturb}\,,
\end{align}
where
\begin{equation}\label{eq:chi_iso}
   \chi^{(0)}(T) = \lim_{h\to 0}\partial_h m(T,h,0)\,
\end{equation}
is the zero-field susceptibility of the unperturbed, isotropic
system. The corrections to Eqs.~\eqref{eq:magnetization_perturb}
and \eqref{eq:susceptibility_perturb} are of order $\mathcal{O}(\lambda^2)$
and $\mathcal{O}(\lambda^2 J^2/(k_B T)^2)$, the latter meaning
that temperatures are restricted to the regime $T\gg \lambda J/k_B$.
Thus, if the temperature dependence of the zero-field susceptibility
of the unperturbed isotropic model is known and if the expectation
values $\langle V \rangle_{T,h,0}$ of the perturbation term with
respect to the unperturbed Hamiltonian $\mathcal{H}_0+\mathcal{H}_Z$
can be computed, Eq.~\eqref{eq:susceptibility_perturb} provides a useful
means to determine the anisotropy parameter $\lambda$.

Naively one might try to proceed by measuring the susceptibility of
the full system, i.e.~the left hand side of Eq.~\eqref{eq:susceptibility_perturb},
and fitting the measured data with the computed right hand side,
using $\lambda$ as a fit parameter. One problem with such kind of
procedure would be that offsets and proportionality factors (like
geometry factors or $g$-factors) of the measured susceptibility
$\chi(T)$ are usually unknown. The energy scale $J$ may be unknown
as well, whereas theoretical predictions of $J\chi^{(0)}(T)$ and
$J\partial_h^2\langle V \rangle_{T,h,0}$ are typically functions of
$k_B T/J$. In the literature an estimate of $J$ is sometimes obtained
by fitting the susceptibility $\chi^{(0)}(T)$ of the isotropic
Hamiltonian to measured susceptibility data, neglecting effects of
small anisotropies. This value cannot be used. The coupling $J$
rather has to be extracted, together with offsets, prefactors and the
anisotropy parameter $\lambda$, from the same fit. But if one uses
a single susceptibility curve, the fit can become unstable since
the second term in Eq.~\eqref{eq:susceptibility_perturb} is small
as compared to the first one.

A considerable improvement can be achieved if several
susceptibility curves are recorded with magnetic fields applied in
different directions, say
$\ev^{(i)} = \hat{g}\bm H^{(i)} / |\hat{g}\bm H^{(i)}|$,
$i=1,2,\ldots,n$. For the theoretical analysis we rather rotate the
chain and keep the direction associated to the Zeeman term $\mathcal{H}_Z$ fix. A
rotation does not affect the isotropic, $SU(2)$-invariant part
$\mathcal{H}_0$ of the total Hamiltonian \eqref{eq:Hamiltonian_perturb}, but
transforms the anisotropic part $V$ into generally different
operators $V^{(i)}$. Plugging those into Eq.~\eqref{eq:susceptibility_perturb}
and denoting first order terms by
$\chi^{(i)}_\text{corr}(T) = - \frac{\lambda}{L} \lim_{h\to 0}\partial_h^2 \langle V^{(i)} \rangle_{T,h,0}$
this yields
\begin{align}
  \chi^{(i)}(T) &\simeq \chi^{(0)}(T) + \chi^{(i)}_\text{corr}(T) \label{eq:chi_i_pre}\,.
\end{align}
Taking also the possibility of different offsets $\chi^{(i)}_0$ and
geometry factors $A^{(i)}$ for different directions
$\ev^{(i)}$ into account we arrive at ($i=1,\ldots,n$)
\begin{align}
  \chi^{(i)}(T) &= A^{(i)}\left(\chi^{(0)}(T) + \chi^{(i)}_\text{corr}(T)\right) + \chi_0^{(i)}\,, \label{eq:chi_i}
\end{align}
which is the general form of Eqs.~\eqref{eq:chi_perturbation} of
the main text. Unknown parameters are offsets $\chi_0^{(i)}$,
geometry factors $A^{(i)}$, energy scale $J$ and last but not least
the anisotropy parameter $\lambda$. They can be determined by a
simultaneous fit of all Eqs.~\eqref{eq:chi_i} with $i=1,\ldots,n$
to the measured data. An advantage of a combined fit (instead of
two individual fits) is that it stabilizes the algorithm. The
two correction terms in Eqs.~\eqref{eq:chi1_perturbation} and
\eqref{eq:chi2_perturbation} are in a way counteractive to each other.

If one is merely interested in a rough estimate of $\lambda$ rather
than in all parameters including offsets and geometry factors of
the experimental susceptibility data, and if the isotropic
susceptibility $\chi^{(0)}(T)$ has a maximum at a known temperature
$T_\text{max}^{(0)}$, one can use a simplified procedure which
requires no fitting and only needs two different directions of the
magnetic field, labeled $1$ and $2$ in the following. It further
does not require knowledge of the full temperature dependence
of the isotropic susceptibility $\chi^{(0)}(T)$. If the
perturbation parameter $\lambda$ is small enough the maximum of
$\chi^{(0)}(T)$ at $T_\text{max}^{(0)}$ gets slightly shifted by
$\chi^{(1,2)}_\text{corr}(T)$, resulting in two new maxima at
$T^{(1)}_\text{max}$ and $T^{(2)}_\text{max}$. The difference of
these two temperatures can be computed as
\begin{equation}\label{eq:ratio_Tmax}
  T^{(1)}_\text{max}-T^{(2)}_\text{max}
     \approx - \left.\frac{\partial_T \chi^{(1)}_\text{corr}(T)
         - \partial_T\chi^{(2)}_\text{corr}(T)}
       {\partial_T^2 \chi^{(0)}(T)}
       \right|_{T=T^{(0)}_\text{max}}\,,
\end{equation}
where we used expansions of $T^{(1,2)}_\text{max}$ up to first
order in $\lambda$ and implicit differentiation. Inserting the
definitions of $\chi^{(1,2)}_\text{corr}(T)$ and solving for
$\lambda$ yields
\begin{equation}\label{eq:lambda_from_chi}
  \lambda \approx A_0\frac{T^{(1)}_\text{max}-T^{(2)}_\text{max}}{T^{(0)}_\text{max}}
\end{equation}
with
\begin{equation}\label{eq:A0_prefactor}
  A_0 = \left.\frac{T \partial_T^2 \partial_h \langle S^z\rangle_{T,h,0}}
                 {\partial_T\partial_h^2\left(\langle V^{(1)} \rangle_{T,h,0}
          - \langle V^{(2)} \rangle_{T,h,0}\right)}
          \right|_{\substack{\hspace{-3ex} h=0 \\ T=T^{(0)}_\text{max}}}\,.
\end{equation}
A particular example where $A_0$ and $T^{(0)}_\text{max}$ can be
explicitly calculated is presented below.

\subsection{\label{app:susceptibility_example} Example: Spin-1/2 XXZ chain}

For a single spin-1/2 XXZ chain, Eq.~\eqref{eq:XXZ-model} with 
aniso\-tropy/perturbation parameter $\delta=\lambda$, one can measure 
the zero-field susceptibility in a magnetic field parallel to the 
anisotropy axis ($\chi^{(\|)}$) and perpendicular to it ($\chi^{(\perp)}$). 
After a suitable spin rotation, which brings the Zeeman term to the 
form $-h S^z$, the corresponding perturbations become
$V^{(\|)} = J\sum_{j}s_j^z s_{j+1}^z$ and $V^{(\perp)} =
J\sum_{j}s_j^x s_{j+1}^x$. Their expectation values can be computed
exactly by solving non-linear integral equations which arise in
the context of the quantum transfer matrix approach to the thermodynamics
of integrable models (see comment between Eqs.~\eqref{eq:chi2_corr}
and \eqref{eq:chi_perturbation} of the main text). Inserting the
expectation values of $V^{(\perp)}$ and $V^{(\|)}$ into
Eq.~\eqref{eq:A0_prefactor} we obtain
\begin{equation}\label{eq:A0_prefactor_XXZ}
  A_0 = \left.\frac{T \partial_T^2 \partial_h \langle s_1^z\rangle_{T,h,0}}
                 {J\partial_T\partial_h^2\langle s_1^z s_2^z - s_1^x s_2^x\rangle_{T,h,0}}
          \right|_{\substack{\hspace{-3ex} h=0 \\ T=T^{(0)}_\text{max}}} \approx 2.39\,,
\end{equation}
where we used $T^{(0)}_\text{max} \approx 0.64085 J/k_B$.\cite{Johnston2000} Therefore,
\begin{equation}\label{eq:delta_from_chi}
  \lambda \approx 2.39\,\frac{T^{(\|)}_\text{max}-T^{(\perp)}_\text{max}}{T^{(0)}_\text{max}} \approx 2.39\left(\frac{T^{(\|)}_\text{max}}{T^{(\perp)}_\text{max}}-1\right)  \,.
\end{equation}

Note that in our notation used in the analysis of the compound CPB (see 
main body of the text, in particular Eqs.~\eqref{eq:chi1_corr} and 
\eqref{eq:chi2_corr} in Sec.~\ref{sec:magnetization}) the labels 
$(\|)$ and $(\perp)$ mean parallel and perpendicular to the 
crystallographic $c$ axis rather than to the anisotropy axis. Due 
to the special arrangement of the anisotropy axes in CPB, the difference of 
the two perturbation terms is minus one half of the difference of 
the two perturbation terms of the single anisotropic chain. Hence, 
the prefactor $A_0$ is twice as big and negative, $A_0 \approx -4.78$, 
and the formula to estimate $\delta$ from the positions of the 
maxima reads
\begin{equation}\label{eq:delta_from_chi_CPB}
  \delta \approx -4.78\left(\frac{T^{(\|)}_\text{max}}{T^{(\perp)}_\text{max}}-1\right)  \,.
\end{equation}
A simple fit of our CPB data around the locations of the maxima, 
$T\in [25\,\text{K},42\,\text{K}]$, yields $T^{(\|)}_\text{max} \approx 33.25\,\text{K}$, $T^{(\perp)}_\text{max}
\approx 33.05\,\text{K}$, $T^{(\|)}_\text{max}/T^{(\perp)}_\text{max}
\approx 1.006$ and hence $\delta \approx -0.03$. Note that
``around the locations of the maxima'' is ambiguous and that outliers and
asymmetry of the maxima caused some difficulties. We
determined an optimal temperature range using a polynomial of degree three as fit function.

The value of the isotropic exchange interaction can as well
be estimated from the temperature value at the maximum
$T^{(\|)}_\text{max}$ using the theoretical prediction
$J/k_B \approx T^{(\|)}_\text{max}/(0.64085+\delta/8+\delta^2/20)
\approx 52.2\,\text{K}$. The $\delta$-corrections in the
denominator are obtained by varying $\delta$, calculating for
each $\delta$ the quantity $k_B T^{(\|)}_\text{max}/J$ exactly, 
i.e.~to high numerical precision by solving non-linear integral
equations,\cite{Kluemper1992, Johnston2000} and approximating
the resulting curve around $\delta=0$ by a polynomial of degree
two. For $\delta=0$ the exact result of Ref.~\onlinecite{Johnston2000}
is reproduced.

The values of $J$ and $\delta$ obtained by this simplified
procedure are compatible with the values $\delta = -0.019$ and
$J/k_B=52.0\,\text{K}$ obtained by a fit to the data over almost
the entire temperature range (see Sec.~\ref{sec:magnetization}),
omitting only very low temperatures, where the perturbation
expansion is not valid, and temperatures above room temperature,
where the susceptibility data are less precise.

\section{\label{app:ESR_parameters} Theoretical description of ESR parameters}

In this section we discuss the ESR parameters `resonance shift' and
`linewidth' and derive some expressions used in the main body of the
manuscript. We consider an interacting spin system in a homogeneous
magnetic field (in $z$ direction) which couples to the total spin. If
the interactions between the spins are purely isotropic, like e.g.\
in Eq.~\eqref{eq:XXZ-model} with $\delta=0$, the Hamiltonian of the
spin system commutes with the total spin, and the dynamical
susceptibility $\chi''_{+-}(\omega,h)$ of Eq.~\eqref{eq:dynamical_susceptibility}
simplifies to $\chi''(\omega,h) = \pi m(T,h)\,\delta(\omega-h/\hbar)$, where
$m(T,h) = \langle s_1^z \rangle_{T,h,0}$ is the magnetization per lattice
site. This means that the absorbed intensity has a single sharp resonance
peak at the paramagnetic resonance frequency $\omega=g\mu_B\mu_0 H/\hbar$. 
The total magnetic moment of the spin system rotates about the magnetic 
field direction. In a weakly anisotropic system, e.g.\ Eq.~\eqref{eq:XXZ-model} 
with $\delta\neq 0$, energy is transferred from the rotation to internal
excitations which causes a shift and a broadening of the paramagnetic
resonance peak.

We wish to identify appropriate measures for this `resonance
shift' and `linewidth', that are both, accessible by theory and
extractable from experimental data. In the introduction of this
manuscript we have discussed different measures of these ESR
parameters, in the first place the maximum position of the peak 
and its width at half height. The latter can be easily read off 
from measured absorption curves, but are unfortunately so far 
inaccessible by theory. A measure which is more convenient for a 
theoretical description is defined in terms of moments of the 
absorption profile $I(\omega,h)/I_0$ (`method of moments'; see 
Sec.~\ref{sec:general_remarks}). It requires an integration over all frequencies or 
fields. This, in turn, is often problematic with experimental data, 
since absorption profiles away from a close vicinity of the location 
of the maximum of the peak may be heavily distorted by systematic 
and statistical errors like underground, noise, and drift.

A problem with the two different kinds of measures discussed above
is that they may behave quite differently as functions of temperature
and magnetic field and therefore cannot be compared naively.
For instance, if we consider experimental data of the width at half
height as function of temperature (see Fig.~\ref{fig:CPX_X-Band_T-dep})
and theoretical predictions for the linewidth of $I(\omega,h)$
based on its moments (see e.g.~Refs.~\onlinecite{Brockmann2011,Brockmann2012}),
they show different monotonic behavior, and we observe a clear mismatch, in
particular at low temperatures. We have discussed two possible
explanations of this discrepancy in the introduction:
distributions with `heavy tails' and different `directions' used
in theory (`$\omega$-direction', i.e.~fixed magnetic field) and
in experiments (`$h$-direction', i.e.~fixed frequency).

In order to address the first problem we shall suppress the spectral
weight of the frequency tails by considering the so-called shape function
$8I(\omega,h)/[\omega(1-e^{-\hbar\omega/(k_B T)})]$ instead of the absorbed 
intensity. For the XXZ spin chain with anisotropy axis parallel
to the magnetic field it becomes a function of the difference
$\hbar\omega-h$ in the high-temperature limit.\cite{Brockmann2012}
Hence, in this limit, the second problem, the inequivalence of $h$-
and $\omega$-directions, is resolved as well. We also believe that
$\omega$- and $h$-directions remain more or less comparable at
infinite temperature even if the anisotropy axis of the XXZ spin
chain is tilted away from the direction of the magnetic field.
Therefore, at high temperature, using the shape function should
reduce both causes for the mismatch (`heavy tails' and `different
directions') at the same time.

At lower temperatures the situation is different. Although the
spectral weight of the high-frequency tails of the normalized
shape function is suppressed as compared to the weight in the tails
of the absorbed intensity $I(\omega,h)/I_0$ (by a factor
$\omega(1-e^{-\hbar\omega/(k_B T)})$), there is still a mismatch
between experimental data and the moment-based linewidth of the
numerically computed shape function.\cite{Brockmann2012}
Our numerical investigations have shown that the linewidth as
obtained from an integration in $h$-direction has the
same monotonic behavior as the width at half height (at least for
temperatures $T\geq J/k_B$), but a quantitative discrepancy remains
between experimental linewidth, measured as width at half height,
and moment-based linewidths calculated by means of the shape
function.

In order to resolve this discrepancy one either has to find a way to
enhance the quality of the experimental data, rendering computations
of moments of the absorption line possible, or one has to find
other theoretical measures for the `resonance position' and `linewidth'
that can be computed and have a known relationship to the experimentally
accessible measures `maximum position' and `width at half height'.
Below we generalize the moment-based approach by introducing
certain `cut-off functions' which effectively restrict the range
of integration in the definition of the moments to a vicinity
of the maximal absorption and suppress the experimentally inaccessible
high-frequency tails.

At first sight this may look like a simple remedy to the
above described problems. However, on the theoretical side,
a cut-off in general spoils our method to calculate the
moments. Moments are relatively easy to calculate if the
cut-off is sent to infinity and if we consider the dynamical
susceptibility rather than the shape function. In this
case, the moments can be expressed in terms of certain
static short-range correlation functions.\cite{Brockmann2011,
Brockmann2012} Still, as we shall point out below, moments
of a shape function restricted by a cut-off can be analyzed,
if we take into account the simplifications coming from a
perturbation theory in small anisotropy parameter $\delta$
and from a high-temperature expansion.

After presenting the precise definition and some properties of
moments in the next subsection, we will focus on
two cases. In the first case we keep the angle between
magnetic field and anisotropy axis constant, $\vartheta = 90^\circ$,
and exploit only the smallness of the anisotropy. The zeroth and
first moment turn out to be independent of the cut-off to lowest
order in $\delta$. This means that they provide a measure for the
resonance shift for all temperatures that is compatible with
the experimentally determined peak position. For the second
moment we have to resort to a numerical calculation of certain
time-dependent correlation functions. In 3d systems the decay of 
these functions is fast (within a time scale of order $\hbar/J$)
which leads to Lorentzian-like spectra with linewidths proportional
to $\delta^2(1+\cos^2\vartheta)$. In 1d systems they decay much 
slower (usually algebraically as $t^{-\gamma}$), leading to possibly 
differently shaped spectral lines with a broader but still narrow 
central peak.\cite{Hennessy1973} It turns out that the small cut-off 
in frequency, required for the narrow absorption lines we have 
observed in our experiments, would make it necessary to calculate 
these time-dependent correlation functions for long times, which is 
beyond the scope of our numerical method.

In the second case we consider the angular dependence of the
moments in the high-temperature regime $T\gg J/k_B$ and for
small anisotropy parameter $\delta$. Then, the two lowest
moments can be calculated explicitly and determine the angular
dependence of the resonance shift. The cut-off further enables
the analysis of the scaling behavior of the second moment.
This scaling behavior, whose analysis is supported by 
numerical investigations (see below), connects the second moment with the
experimentally determined width at half height. This way
we derive a new formula for the angular dependence of the linewidth, 
Eq.~\eqref{eq:width_eta}, which is consistent with the picture
of `inhibited-exchange narrowing'\cite{Hennessy1973} in one
dimension.

\subsection{\label{app:generalized_moments} Moments of the shape function}

In the following we set $k_B = \mu_B = \hbar = 1$.
We will restore the standard units at the very end by
dimensionality considerations. The function
\begin{equation}\label{eq:shape_function2}
  f_{\alpha\beta}(\omega,h) = \frac{2}{L}\int_{-\infty}^{\infty}{\rm d} t \:
     e^{i\omega t}\left\langle S^\alpha(t) S^\beta\right\rangle_{T,h,\delta}
\end{equation}
is called the shape function (of a chain of length $L$). The
difference to the dynamical susceptibility is just a multiplicative
factor $(1-e^{-\omega/T})/4$. This can be seen by expressing the
function $f_{\alpha\beta}$ by its Lehmann series, as shown
e.g.~in Appendices A.7, 8 of Ref.~\onlinecite{Brockmann2012}. The
index pair $(\alpha,\beta)$ takes values $(x,x)$, $(+,-)$, and so
on, depending on the polarization of the incident wave. The time
evolution $S^\alpha(t)=e^{i\mathcal{H}t}S^\alpha e^{-i\mathcal{H}t}$ is governed by
the Zeeman term $-hS^z$ of a magnetic field in $z$ direction plus
the Hamiltonian of the XXZ model, where, in general, the direction
of the anisotropy axis is different from the magnetic field
direction. The full Hamiltonian reads
\begin{subequations} \label{eq:full_Hamiltonian}
\begin{align}
  \mathcal{H} &= \mathcal{H}_\text{xxx} - hS^z + \delta\cdot \mathcal{H}'(\vartheta,\varphi)\,,\\
  \mathcal{H}_\text{xxx} &= J\sum_{j=1}^L(s_j^x s_{j+1}^x + s_j^y s_{j+1}^y+ s_j^z s_{j+1}^z)\,,\\
  \mathcal{H}'(\vartheta,\varphi) &= J \sum_{j=1}^L [\cos\vartheta\,s_j^z -\frac{\sin\vartheta}{2}(e^{i\varphi}s_j^+ + e^{-i\varphi}s_j^-)] \notag\\
  & \times[\cos\vartheta\,s_{j+1}^z -\frac{\sin\vartheta}{2}(e^{i\varphi}s_{j+1}^+ + e^{-i\varphi}s_{j+1}^-)]\,,\label{eq:Hprime}
\end{align}
\end{subequations}
where $\vartheta$ and $\varphi$ are azimuth and polar angles in the
reference frame $(x,y,z)$. Thermal averages in Eq.~\eqref{eq:shape_function2}
are defined by $\langle A \rangle_{T,h,\delta} = \text{Tr}\{e^{-\mathcal{H}/T} A\} / \text{Tr}\{e^{-\mathcal{H}/T}\}$.

Our aim is now to define theoretical measures of the resonance
shift and the linewidth of the central peak around $\omega = h$
of the absorbed intensity $I(\omega,h)$, which can be connected
to experimental measures of the ESR parameters. In the case of small
anisotropy $\delta$ and not too small applied frequencies $\nu = \omega/2\pi$
of the incident micro waves we may assume that the width of this
peak is small compared to the resonance field close to $h=\omega$.
This assumption has several important consequences.
\begin{enumerate}
    \item First of all, it is reasonable to assume that, when
    investigating only the central peak (in a proper definition of
    ESR parameters), it does not matter whether we consider the
    shape function $f(\omega,h)$ as function of frequency $\omega$
    for fixed field $h$ (`$\omega$-direction') or the other way
    round (`$h$-direction'). In this section we will focus on the
    former set-up.
    \item Secondly, additional factors like $\omega/2$ (see paragraph
    below Eq.~\eqref{eq:dynamical_susceptibility} of the main text)
    and $(1-e^{-\omega/T})/4$, which connect the shape function
    $f_{\alpha\beta}$ with the intensity $I$, can be neglected since
    they are almost constant over the whole region in which $I(\omega,h)$
    is non-negligible. Hence, ESR parameters of the shape function
    should be comparable to those of $I$ (i.e.~equal up to leading order),
    as long as tails of $f_{\alpha\beta}$ are not taken into
    account in their definition.
    \item A third consequence is that for a linearly polarized
    incident wave we can neglect all terms in the expansion
    $f_{xx} = \frac{1}{4}(f_{++} + f_{+-} + f_{-+} + f_{--})$
    except for $f_{+-}$. The neglected terms either belong to the
    peak around $\omega = -h$ and are therefore small for $\omega=h$
    ($f_{-+}$) or are very small for all frequencies ($f_{\pm\pm} \ll f_{\pm\mp}$).
    \item Last but not least, due to the previous point, the
    leading orders of the ESR parameters do not depend on $\varphi$.
    Hence, we may choose any value of $\varphi$ in Eq.~\eqref{eq:Hprime},
    e.g.~$\varphi=0$ for convenience.
\end{enumerate}
In conclusion, we may focus on the peak around $\omega = h$ of
the shape function $f_{+-}$. We omit the index $+-$ and denote
it by
\begin{equation}\label{eq:shape_function}
  f(\omega) = \frac{2}{L}\int_{-\infty}^{\infty}{\rm d} t \:
     e^{i(\omega-h) t}\left\langle e^{iht}S^+(t) S^-\right\rangle_{T,h,\delta}\,.
\end{equation}

In order to analyze the corresponding resonance shift and linewidth
we define the shifted moments
\begin{equation}\label{eq:generalized_moments_def}
  m_n(\Omega) = \int_{-\infty}^{\infty}\frac{{\rm d}\omega}{2\pi}\:
     \mu_{\Omega}(\omega-h) (\omega-h)^n f(\omega)\,.
\end{equation}
This definition differs from the one of Refs.~\onlinecite{Brockmann2011,Brockmann2012}
in that we have inserted a cut-off function $\mu_{\Omega}$ under the integral.
As discussed in the introduction of this section, this function is
supposed to suppress the high-frequency tails of the shape function
that are invisible in the experiments. We imagine $\mu_{\Omega}$ as a
symmetric function which falls off rapidly for large arguments and
depends on a cut-off $\Omega$. To keep notations simple, we drop 
in the following the cut-off dependence of $\mu_{\Omega}$ and 
$m_n(\Omega)$. We consider $\mu$ together with its Fourier transform 
$\hat{\mu}(t) = \int_{-\infty}^\infty\frac{{\rm d}\omega}{2\pi}\:
\mu(\omega)e^{i\omega t}$. The precise form of these functions does
not matter for our arguments below. Examples are
\begin{subequations}\label{eq:mu}
\begin{align} \label{eq:mu_chi}
  \mu(\omega) &= \chi_{[-\Omega,\Omega]}(\omega) \,,
     &\hat{\mu}(t) &= \frac{\sin(\Omega t)}{\pi t}\,,\\[0.3ex] \label{eq:mu_sin}
  \mu(\omega) &= \frac{\sin(\omega/\Omega)}{\omega/\Omega} \,,
     &\hat{\mu}(t) &= \frac{\Omega}{2}\chi_{[-\frac{1}{\Omega},\frac{1}{\Omega}]}(t)\,,\\[0.3ex]
  \mu(\omega) &= e^{-\frac{\omega^2}{2\Omega^2}} \,,
     &\hat{\mu}(t) &= \frac{\Omega}{\sqrt{2\pi}}e^{-\frac{t^2\Omega^2}{2}}\,.
     \label{eq:mu_gauss}
\end{align}
\end{subequations}
Here, $\chi_I$ is the characteristic function of the interval $I$.
In general, we require that the cut-off function depends on
a cut-off $\Omega$ in such a way that $\lim_{\Omega
\rightarrow \infty} \mu(\omega) = 1$.

Using the Fourier transform $\hat{\mu}(t)$, the moments can be expressed as
\begin{equation}\label{eq:generalized_moments_t_integral}
  m_n = \int_{-\infty}^{\infty}{\rm d}t\: \hat{\mu}(t)(i\partial_t)^n\hat{f}(t)\,,
\end{equation}
where $\hat{f}(t) = \frac{2}{L}\left\langle e^{iht}S^\alpha(t) S^\beta\right\rangle_{T,h,\delta}$\,.
For the lowest moments we thus obtain the following explicit expressions,
\begin{align}
  m_0 &= \frac{2}{L}\int_{-\infty}^\infty {\rm d}t\:\hat{\mu}(t)\left\langle e^{iht}S^+(t) S^-\right\rangle_{T,h,\delta}\,,\\[0.3ex]
  m_1 &= \frac{2\delta}{L}\int_{-\infty}^\infty {\rm d}t\:\hat{\mu}(t)\left\langle e^{iht}[S^+,\mathcal{H}'](t) S^-\right\rangle_{T,h,\delta}\,,\\[0.3ex]
  m_2 &= \frac{2\delta^2}{L}\int_{-\infty}^\infty {\rm d}t\:\hat{\mu}(t)\left\langle e^{iht}[S^+,\mathcal{H}'](t) [\mathcal{H}',S^-]\right\rangle_{T,h,\delta}\,.
\end{align}

The shape function is real and positive. If we divide by the zeroth 
moment with infinite cut-off, $m_0^{(\infty)}=\lim_{\Omega\to\infty}m_n(\Omega)$, 
its integral over all frequencies is normalized to one. We may 
therefore interpret $f(\omega)/m_0^{(\infty)}$ as a distribution 
function. If this function has a single symmetric peak, then the 
position of its maximum agrees with the average frequency 
$\langle \omega \rangle$, and the first moment becomes 
a measure for the resonance shift,
\begin{equation} \label{eq:averagef}
     s = \langle \omega - h \rangle =  m_1^{(\infty)}/m_0^{(\infty)}\,.
\end{equation}
Similarly, the variance
\begin{equation} \label{eq:variancef}
     \Delta \omega = \sqrt{m_2^{(\infty)}/m_0^{(\infty)} - s^2}
\end{equation}
may be considered as a measure for the width of the peak.
Such an interpretation of the variance is common within
the context of Heisenberg's uncertainty relation. Still, if the
distribution function is not just a Gaussian, the variance and the
more intuitive width at half height may assume rather different
values. For this reason we cannot directly compare the width
calculated by means of Eq.~\eqref{eq:variancef} with linewidths as
usually obtained in ESR experiments. 
The only remaining question is if we can determine the moments 
$m_n(\Omega)$ theoretically for small cut-off $\Omega$, which we 
address in the following subsections.

\subsection{Small anisotropy}

The expressions for the moments simplify considerably if we expand
them for small $\delta$ around the isotropic point,
\begin{subequations}\label{eq:mn_tilde}
\begin{align}
  m_0 &\simeq \frac{2}{L}\left\langle S^+ S^-\right\rangle_{T,h,0} + \mathcal{O}(\delta)\,, \label{eq:m0_tilde}\\[0.3ex]
  m_1 &\simeq \frac{2\delta}{L}\left\langle [S^+,\mathcal{H}'] S^-\right\rangle_{T,h,\delta} + \mathcal{O}(\delta^3)\notag\\
  &\quad + \frac{2\delta^2 J}{iL}\int_{-\infty}^\infty {\rm d}t\:\hat{\mu}(t)\int_{0}^t {\rm d}t_1\: e^{iht_1}\notag\\&\qquad\qquad\qquad \left\langle [S^+,\mathcal{H}'](t_1) [\mathcal{H}',S^-]\right\rangle_{T,h,0}^{(0)}\,,\label{eq:m1_tilde} \displaybreak[0] \\[0.3ex]
  m_2 &\simeq \frac{2\delta^2}{L}\int_{-\infty}^\infty {\rm d}t\:\hat{\mu}(t)e^{iht}\left\langle [S^+,\mathcal{H}'](t) [\mathcal{H}',S^-]\right\rangle_{T,h,0}^{(0)}\notag\\&\qquad\qquad\qquad + \mathcal{O}(\delta^3)\,. \label{eq:m2_tilde}
\end{align}
\end{subequations}
Here the superscript $(0)$ indicates that the time evolution is
generated by $\mathcal{H}_0 = \mathcal{H}_\text{xxx} - hS^z$. As in case of the anisotropic corrections to the susceptibilities
our derivation guarantees the validity of the above formulae for
temperatures $T \gg \delta J$.

The most striking feature of the moments $m_0$ and $m_1$ in
\eqref{eq:m0_tilde} and \eqref{eq:m1_tilde} is that, to lowest
order in $\delta$, they do not depend on the cut-off. Hence, we may
assume that the cut-off is small. Since our measured resonance
peaks for CPB are moreover rather symmetric, the shift of the
position of the maximum should be well described by
Eq.~\eqref{eq:averagef} for all temperatures $T \gg \delta J$.
Since the resonance shift is robust against changes of the
high-frequency tails that are symmetric with respect
to $\omega-h$ we expect that the validity of Eq.~\eqref{eq:averagef}
extends down to low temperature. For the second moment, on the
other hand, the cut-off dependence remains. Inserting the Hamiltonian
\eqref{eq:full_Hamiltonian} into Eqs.~\eqref{eq:mn_tilde}, 
the leading orders in $\delta$ read
\begin{subequations}\label{eq:mn_tilde_ang_dep}
  \begin{align}
    m_0 & = \frac{4\left\langle s_1^z\right\rangle_{T,h,0}}{1-e^{-h/T}} \,, \label{eq:m0_tilde_ang_dep}\\[0.5ex]
    m_1 & =  (3\cos^2\vartheta-1)\frac{4J\delta\left\langle s_1^x s_2^x - s_1^z s_2^z \right\rangle_{T,h,0}}{1-e^{-h/T}}\,, \label{eq:m1_tilde_ang_dep}\\[0.5ex]
    m_2 &\simeq \frac{2 \delta^2}{L}\int_{-\infty}^\infty {\rm d}t\:\hat{\mu}(t)e^{iht}\left\langle [[S^+,\mathcal{H}'](t) [\mathcal{H}',S^-]]\right\rangle_{T,h,0}^{(0)}\,.
      \label{eq:m2_tilde_ang_dep}
  \end{align}
\end{subequations}

For the comparison with our experimental data we have to recall
that we recorded the temperature dependence of the ESR parameters
for an external field along the c axis, i.e.~perpendicular to the
anisotropy axes of both of the inequivalent chains in our CPB
sample. This situation corresponds to $\vartheta=90^\circ$ and
$\varphi=0$ in Eq.~\eqref{eq:Hprime} for both chains, which
equally contribute to the resonance. With this choice of the
angles, Eqs.~\eqref{eq:m0_tilde}--\eqref{eq:m2_tilde} turn into
\begin{subequations}\label{eq:mn_tilde_perp}
\begin{align}
  m_0 & = \frac{4\left\langle s_1^z\right\rangle_{T,h,0}}{1-e^{-h/T}} \,, \label{eq:m0_tilde_perp}\\[0.7ex]
  m_1 & =  \frac{4J\delta\left\langle s_1^z s_2^z - s_1^x s_2^x \right\rangle_{T,h,0}}{1-e^{-h/T}}\,, \label{eq:m1_tilde_perp}\\[0.7ex]
  m_2 &\simeq \frac{\delta^2}{8L}\int_{-\infty}^\infty {\rm d}t\:\hat{\mu}(t) g_T(t)
     \label{eq:m2_tilde_perp}
\end{align}
\end{subequations}
with
\begin{multline}
 g_T(t) =\\ 4\sum_{j,k=1}^L\left\langle [(s_j^zs_{j+1}^+ + s_j^+ s_{j+1}^z)(t)(s_k^zs_{k+1}^- + s_k^- s_{k+1}^z) ]\right\rangle_{T,h,0}^{(0)}\\
       + (h \leftrightarrow -h)\,.\label{eq:g_T_temp}
\end{multline}
We use Eq.~\eqref{eq:averagef} with \eqref{eq:m0_tilde_perp} and
\eqref{eq:m1_tilde_perp} in Secs.~\ref{sec:HF-ESR_LT} and
\ref{sec:temperature_dependence} in order to determine the field-
and temperature-dependence of the resonance shift. As for the width
we computed the correlation function in Eq.~\eqref{eq:g_T_temp}
numerically and tried to compare the second moment obtained from
the experimental data with the theoretical value predicted by
Eq.~\eqref{eq:m2_tilde_perp}. It turned out that the frequency
cut-off required by the experimental data is too small for our
numerically available resolution.

\subsection{\label{app:generalized_moment_HTE} Small anisotropy, high temperature and low frequency}

Further simplifications occur at high temperature. We consider the
first three moments, Eqs.~\eqref{eq:mn_tilde_ang_dep}. For temperatures
$T \gg J$ we can expand thermal averages $\langle A\rangle_{T,h,\delta}$
in the small parameter $J/T$. This way we obtain entirely explicit
and cut-off independent expressions for the zeroth and first moments 
(neglecting subleading orders in $J/T$),
\begin{subequations}
\begin{align}
  m_0 &\simeq 1\,, \label{eq:HTE_m0_tilde} \\[0.3ex]
  m_1 &\simeq \frac{J\delta}{4T}\left((1-3\cos^2\vartheta)h
       +\frac{J\delta}{2}(1+\cos^2\vartheta)\right)\,. \label{eq:HTE_m1_tilde}
\end{align}
On the other hand, the high-temperature expression for the second
moment,
\begin{align}
  m_2 &\simeq \frac{J^2\delta^2}{4}\int_{-\infty}^\infty {\rm d}t\:
      \hat{\mu}_{\Omega}(t) g_\infty(t)\left[\frac{(1-3\cos^2\vartheta)^2}{2} \right.
      \notag \\ & + \left. 5\sin^2\vartheta\cos^2\vartheta\cos(ht)+
                \frac{\sin^4\vartheta}{2}\cos(2ht)\right] \,, \label{eq:HTE_m2_tilde}
\end{align}
\end{subequations}
remains cut-off dependent and contains the infinite-temperature
dynamical correlation function
\begin{equation}\label{eq:g_infty2}
  g_\infty(t) = \frac{4}{L}\sum_{j,k=1}^L\left\langle e^{i\mathcal{H}_\text{xxx}t}s_j^+s_{j+1}^+
                e^{-i\mathcal{H}_\text{xxx}t} s_k^-s_{k+1}^-\right\rangle_\infty\,,
\end{equation}
where $\langle\cdot\rangle_\infty = \lim_{T\to\infty}\langle\cdot\rangle_{T,h,\delta}
= \text{Tr}(\cdot)/2^L$.

We use Eqs.~\eqref{eq:HTE_m0_tilde} and \eqref{eq:HTE_m1_tilde}
together with Eq.~\eqref{eq:averagef} in order to analyze the angular dependence
of the resonance shift in the high-temperature regime (see
Eqs.~\eqref{eq:res_cond}, \eqref{eq:HTE_shift}, and \eqref{eq:res_cond_2}
in Secs.~\ref{sec:angular_dependence} and \ref{sec:HF-ESR_HT}).

The angular dependence of the linewidth at high temperatures, 
Eq.~\eqref{eq:HTE_linewidth} in Sec.~\ref{sec:angular_dependence},
has been inferred from the scaling behavior of the second moment
that can be calculated from \eqref{eq:HTE_m2_tilde} under certain
assumptions about the size of the cut-off and the asymptotics of
the function~$g_\infty$. The scaling behavior connects the width
at half height with the second moment. The argument proceeds as
follows. The ESR absorption line of our experiments consists of a
single peak located at around $h=\omega$ with a width at half 
height of $2 \eta$. It is reasonable to assume that a rescaling of 
the width $\eta \rightarrow a \eta$, $a > 0$, amounts to a 
rescaling of the shape function
\begin{equation} \label{eq:scaletrans}
     f (\omega+h)\quad \rightarrow\quad \frac{1}{a}\; f\left(\frac{\omega+h}{a}\right) \,.
\end{equation}
This is true, for instance, if the ESR absorption line around the 
location of its maximum is shaped like a Lorentzian 
$f(\omega)=2\eta/(\eta^2+(\omega-h)^2)$.

Under this scaling transformation the second moment 
\eqref{eq:generalized_moments_def} transforms like
\begin{equation}
     m_2 (\Omega) \rightarrow a^2 m_2 (\Omega/a) \,.
\end{equation}
If now $m_2$ is a homogeneous function of degree $\gamma$, then
$m_2 (\Omega) \rightarrow a^{2 - \gamma} m_2 (\Omega)$. It follows
that the ratio $\bigl[m_2 (\Omega)\bigr]^{\frac{1}{2 - \gamma}}/\eta$
is scale invariant and thus
\begin{equation} \label{eq:noscale}
     \eta \propto \bigl[m_2 (\Omega)\bigr]^{\frac{1}{2 - \gamma}} \,,
\end{equation}
which relates the width at half height with the second moment.

\begin{figure}[t]
  \includegraphics[width=\columnwidth]{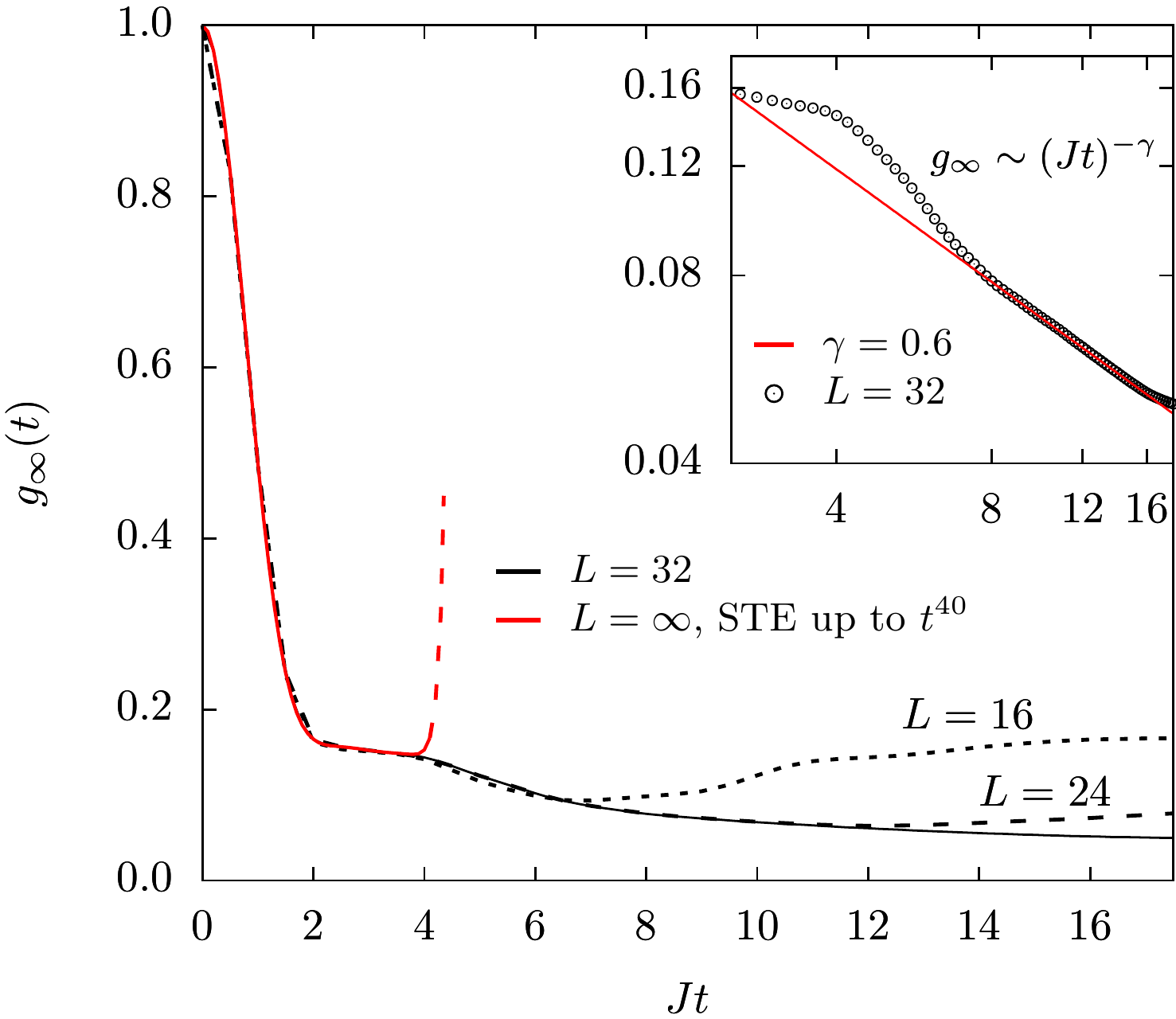}
  \caption{Dynamical correlation function $g_\infty$ as function of
  time $Jt$ for lattice sites $L=16$, $24$, and $32$ (dotted, dashed,
  and solid black lines) together with an exact short time expansion (STE)
  of the infinite chain up to $t^{40}$ (red line). Inset: data for $L=32$
  in a double logarithmic plot (black circles) together with an asymptotic 
  fit (red line) of $g_\infty(t) = \alpha (Jt)^{-\gamma}$ with
  $\gamma=0.6$ and $\alpha=0.272$.}
  \label{fig:g_infty}
\end{figure}

Let us now argue that the second moment is indeed a homogeneous
function of the cut-off if we restrict ourselves to an appropriate
parameter regime. We hypothesize that the function $g_\infty$
in \eqref{eq:g_infty2} behaves for large times, $t\gg 1/J$, as
\begin{equation}\label{eq:g_infty_asymp}
  g_\infty(t) \simeq \alpha (Jt)^{-\gamma}\,,
\end{equation}
where $\alpha$ is of order one and $0<\gamma\leq 1$. We can support
this claim by numerical calculations for finite system sizes up to
$L=32$, as shown in Fig.~\ref{fig:g_infty}. A fit of a straight line to the
double logarithmic data in the asymptotic time regime
$8\leq Jt \leq 16$ provides $\gamma=0.6$ and $\alpha=0.272$ (see
inset of Fig.~\ref{fig:g_infty}). In the main body of the text we
refer to this value as $\gamma_\infty$, to indicate that this is
the value of $\gamma$ at infinite temperature. This value is not in
contradiction to the value $\gamma=0.70$ reported at the end of
Sec.~\ref{sec:angular_dependence}, since this is the measured value
at large but finite temperature ($T/J \approx 6 < \infty$).
We performed further numerical calculations at finite temperatures
(down to $T/J = 1$) which are in accordance with these
findings.

In order to evaluate the time integral in Eq.~\eqref{eq:HTE_m2_tilde}
we choose the cut-off function Eq.~\eqref{eq:mu_sin} for our
convenience. The integral in Eq.~\eqref{eq:HTE_m2_tilde} is then 
restricted to the interval $[- 1/\Omega, 1/\Omega]$. The cut-off
$\Omega$ should not be too small in order to cover the whole
central peak, say $\Omega \gtrsim h$. Such a choice of $\Omega$ is
possible if $h \ll J$, which holds true for our measurements on CPB.
We can therefore neglect all oscillations and approximate
$\cos(ht)\approx \cos(2ht)\approx 1$ in Eq.~\eqref{eq:HTE_m2_tilde}.
We eventually obtain
\begin{equation} \label{eq:HTE_m2_tilde_Omega}
  m_2 (\Omega) \approx \frac{J^2\delta^2(1+\cos^2\vartheta)}{4}
     \frac{2 \alpha}{1 - \gamma} \left(\frac{\Omega}{J}\right)^\gamma
\end{equation}
for the cut-off dependence of the second moment.

Then, Eq.~\eqref{eq:noscale} implies that
\begin{equation}\label{eq:width_eta}
  \eta\, \propto\, J\left[\frac{\delta^2}{4}\left(1+\cos^2\vartheta
                                            \right)\right]^{\frac{1}{2-\gamma}}\,,
\end{equation}
where $\delta$ is the small anisotropy parameter of the Hamiltonian
\eqref{eq:full_Hamiltonian} and $\vartheta$ the angle between
magnetic field and anisotropy axis. Restoring standard units and
assuming that the proportionality factor does not depend on
$\vartheta$ or $\delta$, Eq.~\eqref{eq:width_eta} turns into
Eq.~\eqref{eq:HTE_linewidth} of the main text.

\section{\label{app:spin_structure} Spin configuration at zero temperature}

In this section we present arguments on why the spin structure of the
zero-temperature ordered ground state in CPB is as indicated by the 
arrows in Fig.~\ref{fig:CPB_structure}, in particular why an alignment 
of the assigned magnetic moments along the direction of the chains is 
preferred. The reasoning is based on scaling arguments similar to those 
of Ref.~\onlinecite{SKH02}. We interpret the interchain coupling as a
small perturbation and determine the relevance of the corresponding 
operators in the sense of renormalization group theory. To this end, 
we shall calculate the large distance behavior of the correlation 
function of the interchain operators and compare scaling dimensions of 
the different terms with the marginal value of $2$, the latter being
characteristic for the underlying 1+1 dimensional conformal field 
theory.

We consider two anisotropic spin-1/2 Heisenberg chains of type 
\eqref{eq:XXZ-model} with small anisotropy parameter $\delta<0$ and 
with anisotropy axes perpendicular to the chain direction as well as 
perpendicular to each other. Furthermore, the two chains are 
parallel and shifted against each other by half of the lattice 
constant (zig-zag ladder). This configuration is closely linked to 
the structure of the compound CPB (see the main body of the text). 
We assume a small isotropic (antiferromagnetic) interchain 
interaction, $J' \ll J$, whose Hamiltonian reads
\begin{equation}\label{eq:ici_ham}
  \mathcal{H}_{\rm int} = J'\sum_{j} h_j\,,\quad h_j = \bm s_{j}^{(1)} \cdot \bm s_{j}^{(2)} + \bm s_{j}^{(1)} \cdot \bm s_{j+1}^{(2)}\,,
\end{equation}
where the superscripts $(1)$ and $(2)$ distinguish the two chains. 
Note that for classical systems the geometrical frustration results 
in the vanishing of the interchain coupling for antiferromagnetically
ordered chains. This is not the case for quantum chains. Still, the 
frustration renders the interchain coupling being a perturbation 
close to marginal.

Since $\delta$ is negative each single chain is in the antiferromagnetic 
gapless phase. We parameterize the anisotropy as $\delta = \cos\gamma-1$. 
For small values of $|\delta|$, the inverse relation $\gamma = \text{arccos}(1+\delta)$ 
can be approximated by $\gamma \approx\sqrt{2|\delta|}$. The CPB 
value $\delta \approx -0.02$, for instance, yields $\gamma\approx 0.2$. 
From conformal field theory it is known that large distance 
correlation functions of the XXZ Heisenberg chain at zero 
temperature decay as\cite{LuPe75, Haldane81}
\begin{equation}\label{eq:zero_T_correlators}
  \langle s_1^\alpha s_{r+1}^\alpha\rangle \;\sim\; \frac{(-1)^r}{r^{2x_\pm}}\,,\qquad \alpha = x,y,z\,.
\end{equation}
The exponent is $2x_+=(1-\gamma/\pi)^{-1}$ if $\alpha$ 
coincides with the direction of the anisotropy axis, and it is 
$2x_- =1-\gamma/\pi$ if the $\alpha$ direction is perpendicular 
to it.\cite{KWZ93} The quantities $x_\pm$ are called 
scaling dimensions of spin-spin correlation functions. Since 
$\gamma/\pi$ is small, e.g.\ $\gamma/\pi \approx 0.06$ for CPB, we 
can expand the first exponent as $2x_+ = 1+\gamma/\pi + \gamma^2/\pi^2$. 
Therefore, we have $x_+ + x_- = 1 + \gamma^2/(2\pi^2)$. 

Let us fix the direction of the two chains to $z$ and the directions of their
anisotropy axes to $x$ and $y$, respectively.  After a straightforward
calculation we obtain for the correlation function of the interchain operator
$h_j$ in the ground state of decoupled chains
\begin{align}
  \langle h_1 h_{r+1}\rangle &= \sum_{\alpha,\beta} 
  \langle {s_1^{\alpha}}^{(1)}({s_1^{\alpha}}^{(2)}+{s_2^{\alpha}}^{(2)}) {s_{r+1}^{\beta}}^{\!\!\!\!\!\!(1)}({s_{r+1}^{\beta}}^{\!\!\!\!\!\!(2)}+{s_{r+2}^{\beta}}^{\!\!\!\!\!\!(2)}) \rangle\notag\\[1.3ex] 
  &= \!\!\sum_{\alpha=x,y,z} \langle s_1^{\alpha} s_{r+1}^{\alpha} \rangle^{(1)} \langle (s_1^{\alpha} + s_2^{\alpha} )(s_{r+1}^{\alpha} + s_{r+2}^{\alpha})\rangle^{(2)} \notag\\[1.7ex]
  &\sim\; \frac{4(x_+^2 + x_-^2) + 2(x_+ + x_-)}{r^{2(x_+ + x_- +1)}} + \frac{4x_-^2 + 2x_-}{r^{2(2x_- + 1)}} \,.
\end{align}
Here, the superscripts $(1)$ and $(2)$ again refer to spin 
operators acting on the first or on the second spin chain, 
respectively. The first term of the last line stems from the 
$\alpha=x,y$ contributions, where the spin direction is parallel to 
one of the anisotropy axes and perpendicular to the other one. The 
second term is the $\alpha=z$ contribution, where the spin direction 
is perpendicular to both anisotropy axes at the same time. We infer 
that the scaling dimension of the perturbation 
terms with spin direction $\alpha$ perpendicular to the chain is 
$x_+ + x_- + 1 = 2 + \frac{\gamma^2}{2\pi^2}> 2$, whereas it is 
$2x_- + 1 = 2 -\gamma/\pi < 2$ for the perturbation with spin 
direction along the chain. The marginal scaling dimension is 
$2$. Therefore, the $s^z$-$s^z$ term (and only this one) 
represents a relevant perturbation of the critical system. 

For just two weakly coupled chains this relevant perturbation would 
result in dimer order. In case of infinitely many chains (as in the 
compound CPB), however, we like to argue that true long-ranged
antiferromagnetic order in the $s^z$ components of the local spins 
sets in, which can be interpreted as `collinear spins'.

 
\bibliography{literature_spin_chains}


\end{document}